%% file: main.tex
\documentclass[fleqn,usenatbib]{mnras}

\usepackage{mathptmx}
\usepackage[T1]{fontenc}
\usepackage{ae,aecompl}
\usepackage{graphicx}	
\usepackage{amsmath}	
\usepackage{amssymb}	
\usepackage{verbatim} 
\usepackage[english]{babel}
\usepackage[dvipsnames]{xcolor}
\usepackage{natbib}
\usepackage{txfonts}
\usepackage{pifont}
\usepackage{longtable} 
\usepackage{booktabs} 

\title[Photometric study of the young open clusters]{Photometric study of the young open clusters IC 1442, King 21, and Trumpler 7}

\author[J. Maurya et al.]{
Jayanand Maurya$^{1,2}$\thanks{E-mail: jayanand@aries.res.in},
Y. C. Joshi$^{1}$,
A. S. Gour$^{2}$
\\
$^{1}$Aryabhatta Research Institute of Observational Sciences, Nainital-263002, India
\\
$^{2}$School of Studies in Physics and Astrophysics, Pandit Ravishankar Shukla University, Chattisgarh 492 010, India
}

\date{Accepted XXX. Received YYY; in original form ZZZ}

\pubyear{2020}

\begin{document}

\maketitle

\begin{abstract}
 We carried out UBVR$_{c}$I$_{c}$ photometric study of three poorly studied young open clusters IC 1442, King 21, and Trumpler 7 (Tr 7). We obtained 263, 244, and 128 member stars using \textit{Gaia} DR2 proper motions and parallaxes in IC 1442, King 21, and Tr 7, respectively. The reddening, $E(B-V)$, was derived to be 0.54$\pm$0.04, 0.76$\pm$0.06, and 0.38$\pm$0.04 mag for these clusters. The comparison of observed colour-magnitude diagrams (CMDs) with solar metallicity isochrones yields log(Age) = 7.40$\pm$0.30, 7.70$\pm$0.20, and 7.85$\pm$0.25 yr  and corresponding distances 2847$\pm$238, 2622$\pm$156, and 1561$\pm$74 pc for IC 1442, King 21, and Tr 7, respectively. The estimated mass function (MF) slopes are found to be -1.94$\pm$0.18, -1.54$\pm$0.32, and -2.31$\pm$0.29 for IC 1442, King 21, and Tr 7, respectively. The study of MF slopes determined separately in the inner and the outer regions of these clusters gives a steeper slope in outer region which suggests spatial variation in slope and mass segregation in the clusters. We found evidence of mass segregation after dynamical study in these clusters. The obtained relaxation time, T$_{E}$, is 74, 26, and 34 Myr for the clusters IC 1442, King 21, and Tr 7, respectively. The mass segregation in IC 1442 may be caused by early dynamical relaxation. The estimated T$_{E}$ is well below to the ages of King 21 and Tr 7 which indicates that these clusters are dynamically relaxed. 
\end{abstract}

        

\section{Introduction}\label{introduction}

\begin{figure*}  
\vspace{0 cm}
\hbox{ 
\hspace{0.6 cm}  
  \includegraphics[width=6.0 cm, height=6.0 cm]{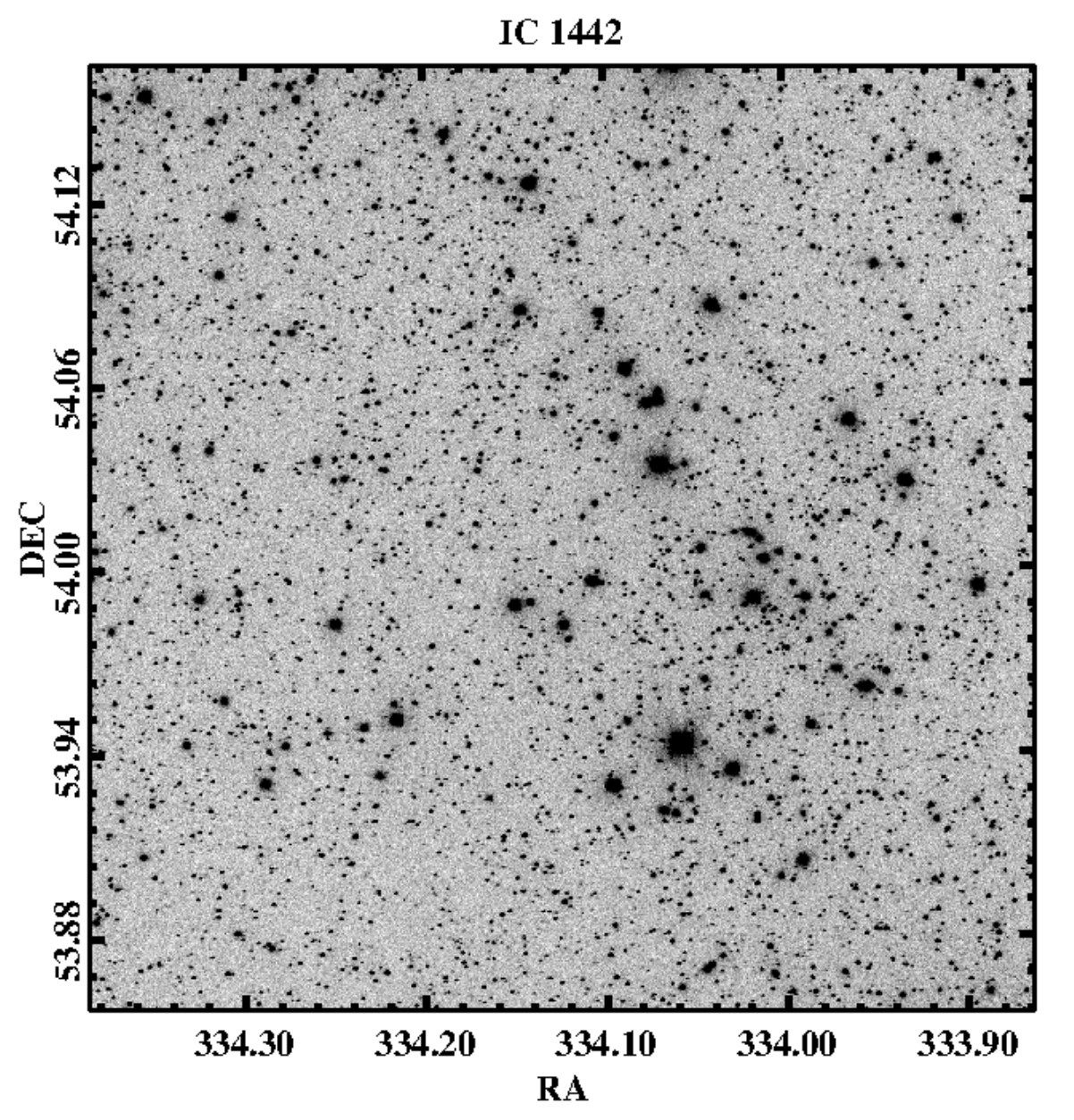}
  \hspace{-0.4 cm} 
  \includegraphics[width=6.0 cm, height=6.0 cm]{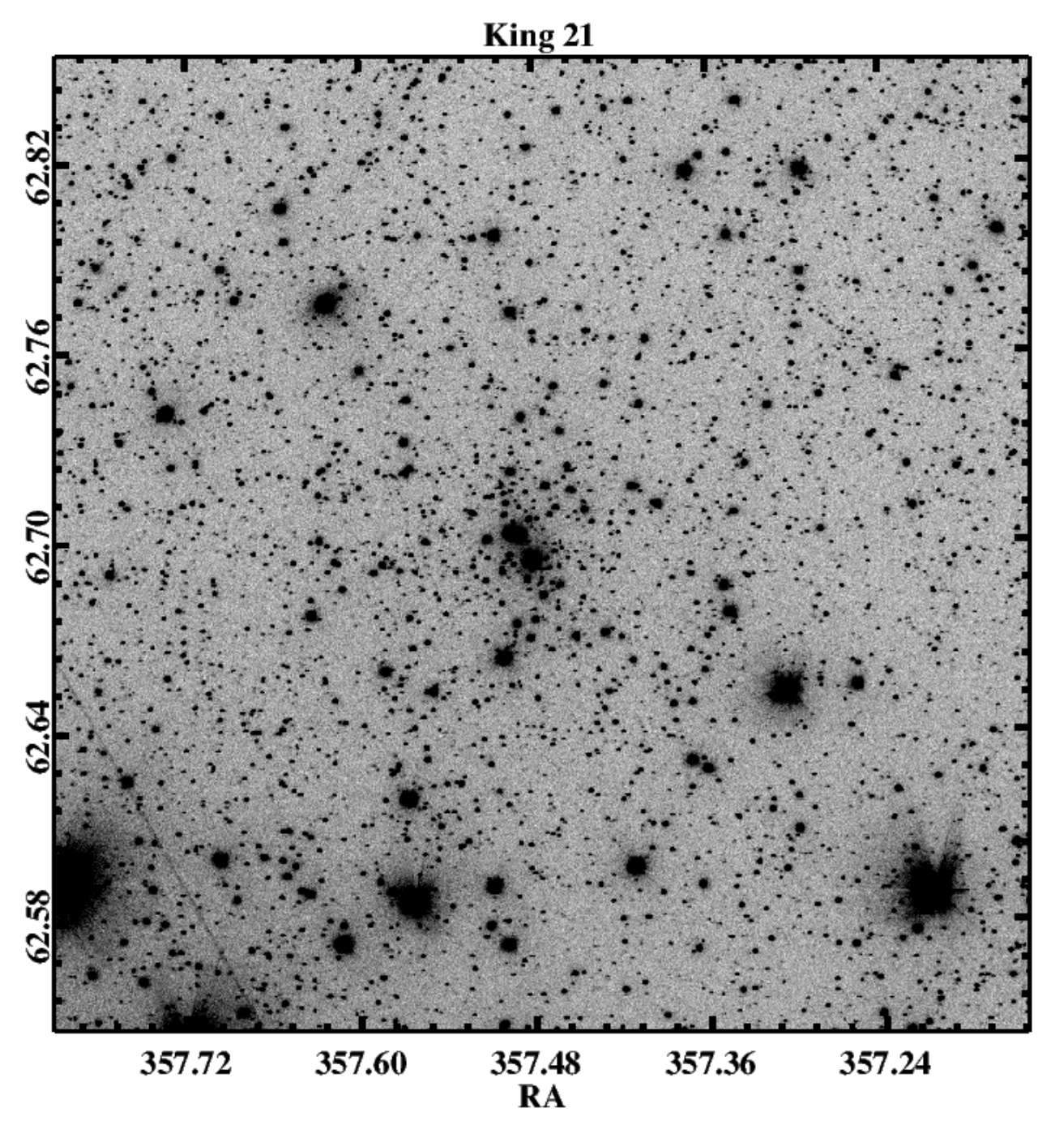}
  \hspace{-0.4 cm}      
  \includegraphics[width=6.0 cm, height=6.0 cm]{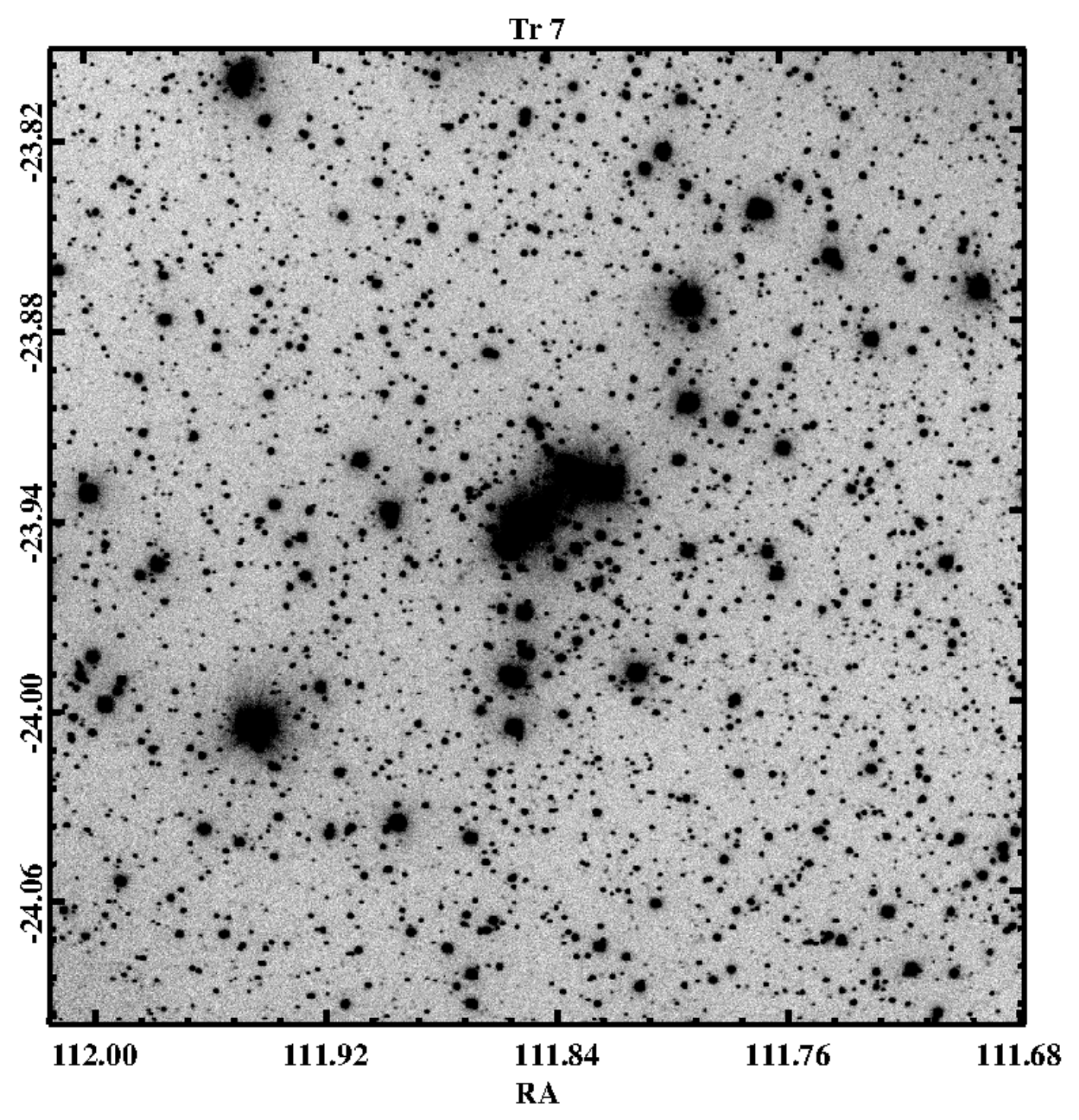}  
  }
\caption{The finding charts for the clusters IC 1442, King 21, and Tr 7. The image sizes are ~$18^{'}$ $\times$ $18^{'}$.} 
  \label{fchart}
\end{figure*}

Star clusters are excellent laboratories to test the theoretical models of stellar evolution as the stars in a cluster are born together in the same molecular cloud therefore they have approximately similar age, distance, and chemical composition. Since open clusters specially young open clusters are mostly found in the relatively dense background of the Galactic disk, it is necessary to separate member stars of the young open clusters from the field stars to accurately determine their physical parameters \citep{2008MNRAS.386.1625C,2018MNRAS.481.3887D}. Before availability of the kinematic data of \textit{Gaia} DR2, membership analysis was available for a very small fraction of known star clusters. However, after availability of \textit{Gaia} DR2 astrometric data with unprecedented accuracy the situation  has changed \citep{2018MNRAS.481.3887D,2019A&A...624A.126C, 2019MNRAS.487.2385M}. The \textit{Gaia} DR2 data provides the better determination of membership of stars which lends an improvement in the estimation of parameters through isochrone fittings \citep{2018MNRAS.481.3887D,2020MNRAS.493.3473A}. A precise membership and physical parameters determination is very useful in the study of the distribution of mass during formation of stars which is known as initial mass function (IMF). The IMF of open clusters help us in having insight in initial condition during star formation and evolution and hence, IMF is considered as fossil record of star formation process in the cluster. Whether the IMF is universal in time and space or it depends upon different star forming conditions is still debated \citep{2010ARA&A..48..339B,2018A&A...614A..43D,2018A&A...620A..39J}.

The young open clusters are important in study of massive stars as they have relatively high number of massive stars which are otherwise not found in intermediate and old open clusters due to stellar evolution. In the mass segregation effect massive stars are mostly concentrated within the central region and low mass stars towards outer region of the cluster. It is still debated whether reason behind mass segregation is dynamical evolution of the cluster or star formation process itself \citep{2018MNRAS.473..849D}. The young open clusters are spread in the spiral arms of the Galactic disk and hence, they are important tools to study the structure and evolution of the Galactic disk \citep{2016A&A...593A.116J,2018MNRAS.480.2386M,2019MNRAS.486.5726D}. The young open clusters are also helpful in study of massive stars and the stellar evolution in the upper part of the Hertzsprung-Russell diagram \citep{2012A&A...544A..64A}. 

We are carrying out a long-term observational program of in-depth photometric study of open clusters at ARIES, Nainital. This program is particularly focused on young and intermediate age open clusters \citep{2012MNRAS.419.2379J, 2014MNRAS.437..804J, 2020MNRAS.492.3602J, 2020MNRAS.494.4713M}. In this paper, we present a comprehensive photometric study of three young open clusters IC 1442, King 21, and Tr 7 as continuation of our ongoing project of investigating open clusters. The cluster IC 1442 is poorly studied cluster and a photometric study is necessary to understand stellar and dynamic evolution of the cluster. The distance to the cluster IC 1442 is reported in \citet{1970A&A.....8..213Y}. The reddening, age, and distance of King 21 are derived using photoelectric data by \citet{1984BASI...12..217M}. The cluster King 21 is included in paper on search of peculiar stars in open clusters by \citet{2007A&A...462..591N}. King 21 is also listed in Be emission study by \citet{2008MNRAS.388.1879M}. There is significant difference in the parameters derived in these studies. A comprehensive photometric study based on membership determined using precise proper motions and parallaxes is helpful in understanding stellar and dynamical evolution in these clusters. The cluster Tr 7 is also a poorly studied young open cluster. The reddening, age and distance modulus of the cluster has been reported in previous studies \citep{1972A&AS....7..133V,2005ApJS..161..118M}. The present study of membership based on proper motions, stellar and dynamical evolution will be useful in study of the cluster properties.
 
This paper is organized as follows. We described observations and data analysis in Section~\ref{observ}. The structural parameters of the clusters are obtained in Section~\ref{RDP}. The stellar membership in the clusters are discussed in Section~\ref{membership} followed by estimation of physical parameters in Section~\ref{param}. The dynamic evolution of the clusters is studied in Section~\ref{dynamic}. We summarized our work in Section~\ref{summary}.

\section{Data}\label{observ}
\subsection{ Observations and calibration}
The observed UBVR$_{c}$I$_{c}$ data used in this study were taken from 1.3-m Devasthal Fast Optical Telescope (DFOT) at Devasthal, India. The telescope is equipped with a 2k$\times$2k CCD camera and have a wide field of view of $\sim 18^{\arcmin}$ $\times$ $18^{\arcmin}$. The CCD gain and read noise are 2 e$^{-}$/ADU and 6.5 e$^{-}$, respectively. We also observed Landolt's standard fields SA 92 and SA 98 at various airmass positions in the sky to obtain atmospheric extinction coefficients on the observed nights. We took long as well as short exposure images to detect both bright and faint stars present in the observed regions. The finding charts for the observed regions of the clusters are given in Figure~\ref{fchart}.

\begin{figure}
\vspace{-0.5cm}
   \centering
  \includegraphics[width=7.0 cm, height=5.0cm]{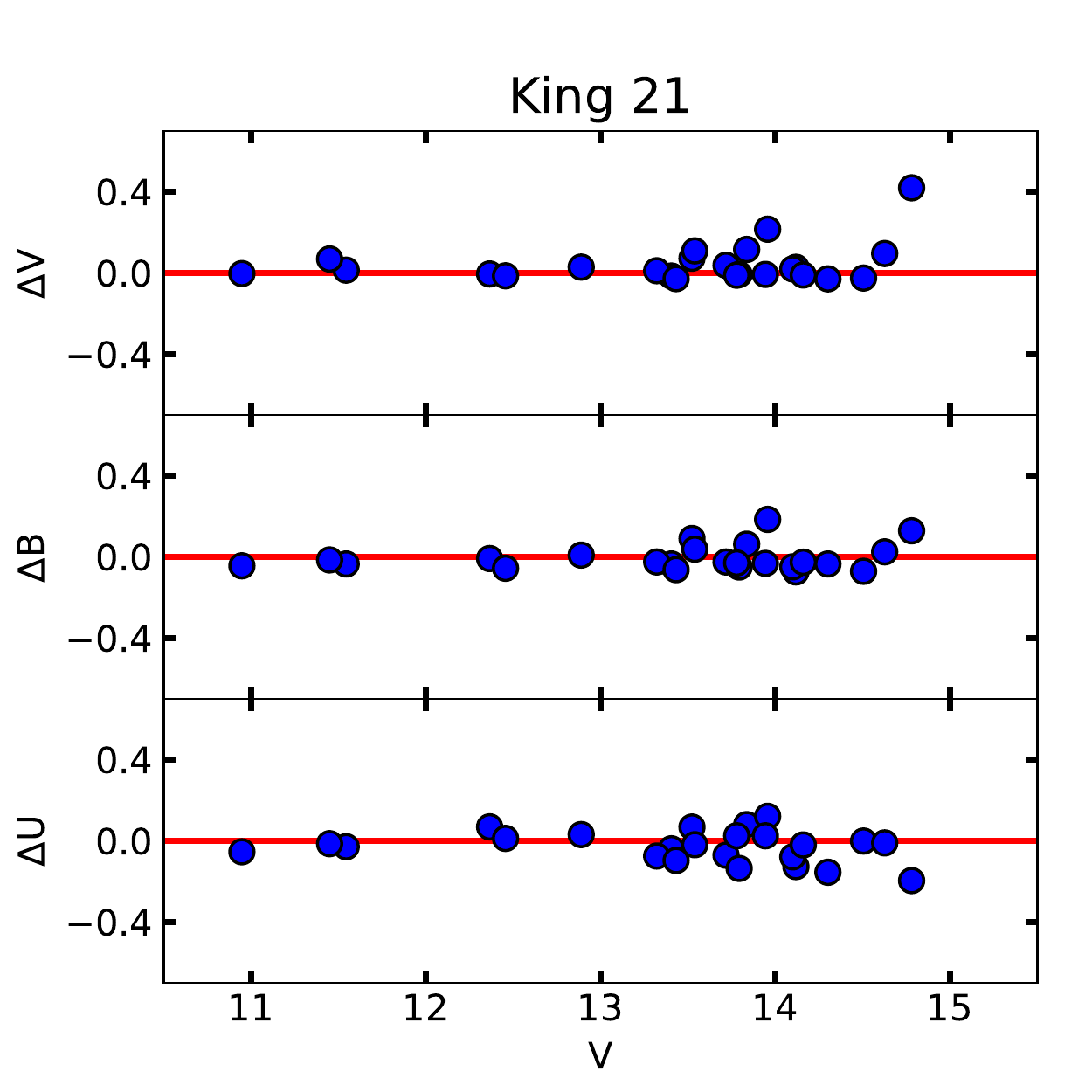}
\caption{Comparison of present photometric data with photoelectric data of \citet{1984BASI...12..217M} for King 21 in U, B, and V bands.} 
 \label{comp}
\end{figure}

The pre-processing like bias subtraction, flat field correction and cosmic ray hits removal of the data was done using the Image Reduction and Analysis Facility (IRAF). We estimated instrumental magnitudes through aperture photometry \citep{1987PASP...99..191S, 1992ASPC...25..297S} and PSF techniques using the DAOPHOT II software. The transformation of the instrumental magnitude to the standard system was achieved using the procedure given by \citet{1992ASPC...25..297S}. The transformation of instrumental magnitude to standard magnitude is described in \citet{2020MNRAS.494.4713M}. The obtained mean photometric error in each magnitude bin for the present study are given in Table~\ref{mag_err}. We compared present photometric data for King 21 with photoelectric data of \citet{1984BASI...12..217M} in Figure~\ref{comp} and found a  good agreement between the two photometries. We, however, could not find photometric data for the clusters IC 1442 and Tr 7 to compare with. \\ 
\input{table01}

\subsection{Archived Data}
In the present study, we used near-IR data in the $J$, $H$, and $K_s$ bands from the Two Micron All-Sky Survey (2MASS) archive \citep{2006AJ....131.1163S}. We converted $K_s$ magnitudes to K magnitudes using the relation provided by \citet{2001AJ....121.2851C}. The 2MASS survey mapped the sky in near-IR bands and have data with a signal-to-noise ratio (S/N) greater than 10 reaching as deep as upto 15.8, 15.1, and 14.3 mag in J, H, and K$_{s}$ bands, respectively. The proper motions and parallaxes were taken from the \textit{Gaia} DR2 archive \citep{2018A&A...616A...1G}. The \textit{Gaia} DR2 provides kinematic data with unprecedented precision. The trigonometric parallaxes from \textit{Gaia} DR2 have mean parallax error up to 0.04 mas and 0.7 mas for sources having G magnitudes upto $\leq$ 15 and 20 mag, respectively. The DR2 provides proper motions with uncertainties up to 0.06, 0.2, and 1.2 mas yr$^{-1}$ for the sources having G magnitudes upto $<$ 15, 17, and 20 mag, respectively.

\subsection{Completeness of the observed data}\label{complete}
\begin{figure}
  \includegraphics[width=8.0cm, height=8.0cm]{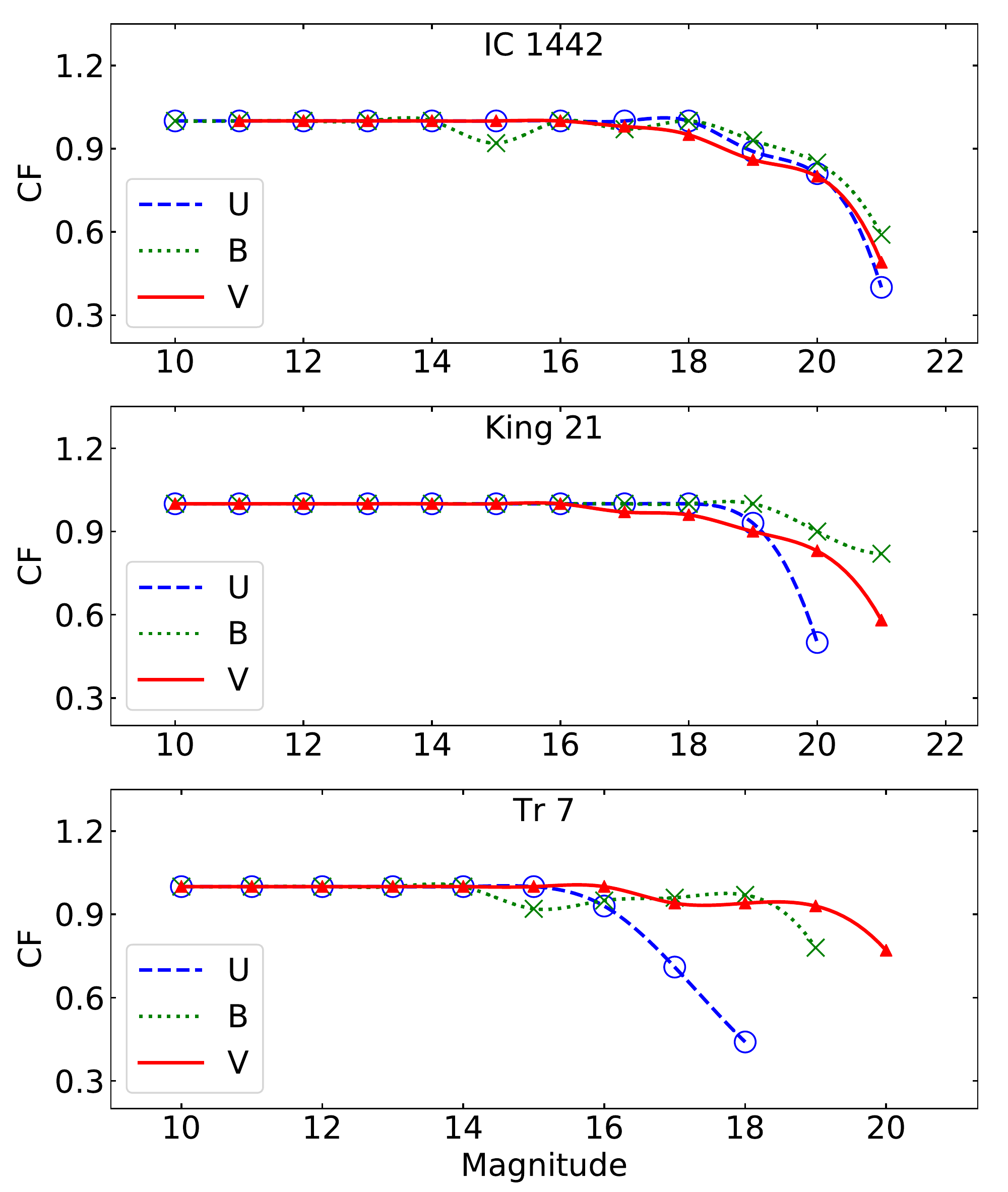} 
\caption{Plots of Completeness Factor as a function of magnitude for IC 1442, King 21 and Tr 7 in U, B, and V bands.}
\label{f_cf}
\end{figure}

It is difficult to detect all stars in the CCD image specially faint stars due to many reasons like crowding of the stars, detection limit etc. The knowledge of compeleteness of the data is thus required to calculate the parameters dependent on the number of stars like luminosity function, mass function, and cumulative distributions of stars with radial distance. In the present study we randomly added artificial stars of different magnitudes in the original images to estimate the completeness of the data. The completeness of the data is quantitatively characterized by the completeness factor (CF). The CF is defined as the ratio of total number of stars recovered to the number of artificial stars added in each magnitude bin. The plots of calculated CF versus magnitude in U, B, and V bands are shown in the Figure~\ref{f_cf}. The completeness decreases towards fainter magnitudes due to increased number of stars in fainter magnitude bins and detection limit of the telescope as can be seen from Figure~\ref{f_cf}. Our data is complete upto 18, 19, and 19 mag in V band for the clusters IC 1442, King 21, and Tr 7, respectively. In this study we used V band data to derive parameters affected by data completeness. 

%
\section{Spatial structure}\label{RDP}
\begin{figure}
   \centering
  \includegraphics[width=8.0 cm,height=9.0 cm]{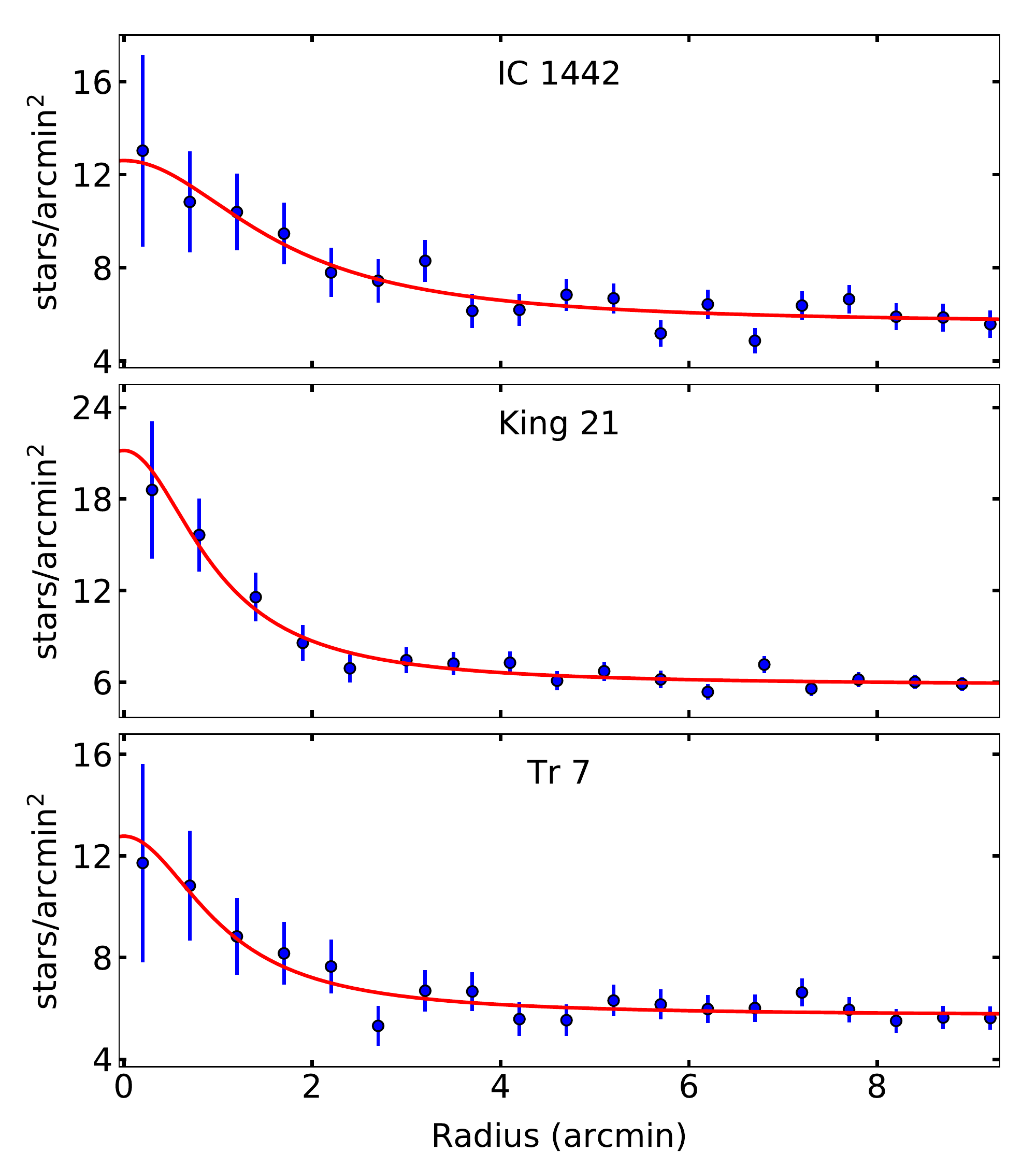}
 \caption{Plots of the stellar density variation with radial distance from cluster center for IC 1442, King 21, and Tr 7. The best fit of radial density profile given by \citet{1962AJ.....67..471K} are presented by red solid curve.}
 \label{rdp}
\end{figure}

The determination of the cluster structure, radial extent, and cluster center is important in study of structural and dynamical evolution of the cluster. The spatial structure of the cluster is also helpful in understanding the distribution of stars within the cluster. However due to irregular shape and sometimes relatively lower density of cluster stars against field stars of the open clusters, the precise determination of cluster center and radial extent becomes difficult \citep{2013A&A...558A..53K}. We estimated cluster center using maximum density method as the center is the point where stellar density is maximum.  The spatial distribution of stars and structure of the clusters were derived using radial density profile (RDP) method. We calculated the number density of stars within concentric annular regions to derive the RDP of the clusters. The width of the annular regions were taken to be $\sim$0$^{\prime}$.5. The larger width of the annular rings are good for determination of sparse open clusters in which fluctuations due to field stars have more impact on determination of cluster extent. The derived cluster center using maximum density method are found to be at celestial coordinates (22:16:03.72, +53:59:29.42), (23:49:52.40, +62:42:15.88), and (07:27:22.86, -23:56:35.87) for the clusters IC 1442, King 21, and Tr 7, respectively. The cluster centers derived by us are in agreement to the centers derived by \citet{2020A&A...633A..99C} as (23:49:54.96, +62:42:18.0) and (07:27:23.76, -23:56:56.4)  for King 21 and Tr 7, respectively. We derived radial density distributions of the clusters by fitting RDP given by \citet{1962AJ.....67..471K} and \citet{1992AcA....42...29K} as following: 
$$
\rho(r) = \rho_{f} + \frac{\rho_{0}}{1 + \left(\frac{r}{r_{c}}\right)^{2}}
$$
where $\rho(r)$, $\rho_{f}$, and $\rho_{0}$ are stellar, field, and central density, respectively. The r$_{c}$ denotes the core radius of the cluster. The core radius is defined as radial distance at which $\rho$(r) becomes half of $\rho_{0}$. The $\chi^{2}$ best fit radial stellar density profiles of the stars detected in V band for the three clusters are shown in Figure~\ref{rdp}. The King profile is presented by solid red curve in the figure. We derived core radii using above relation as 1$^{\prime}$.7$\pm$0$^{\prime}$.5, 1$^{\prime}$.0$\pm$0$^{\prime}$.2, and 1$^{\prime}$.0$\pm$0$^{\prime}$.3 for IC 1442, King 21, and Tr 7, respectively. To derive the cluster radius it is necessary to identify the cluster boundary. We took cluster boundary as a point in radial direction where the radial stellar density is 1$\sigma$ above the field star density. The radial stellar density at the boundary of the cluster can be defined as 
$$
\rho_{lim} = \rho_{f} + \sigma_{f}
$$
\noindent where $\sigma_{f}$ is uncertainty in the radial stellar density $\rho_{f}$. Thus, the cluster radius, r$_{cluster}$, can be calculated as
$$
r_{cluster} = r_{c} \sqrt{\frac{\rho_{0}}{\sigma_{f}} - 1}
$$
\noindent We calculated the cluster radii using above formula on the derived RDPs to be 9$^{\prime}$.5$\pm$0$^{\prime}$.5, 8$^{\prime}$.7$\pm$0$^{\prime}$.5, and 7$^{\prime}$.1$\pm$0$^{\prime}$.5 for the clusters IC 1442, King 21, and Tr 7, respectively. We have summarized the estimated values of the structural parameters in Table~\ref{struc_par}.

\input{table02}

\section{Proper Motion: Membership}\label{membership}
\begin{figure*}
   \hbox{
  \includegraphics[width=17.0 cm,height=10.0 cm]{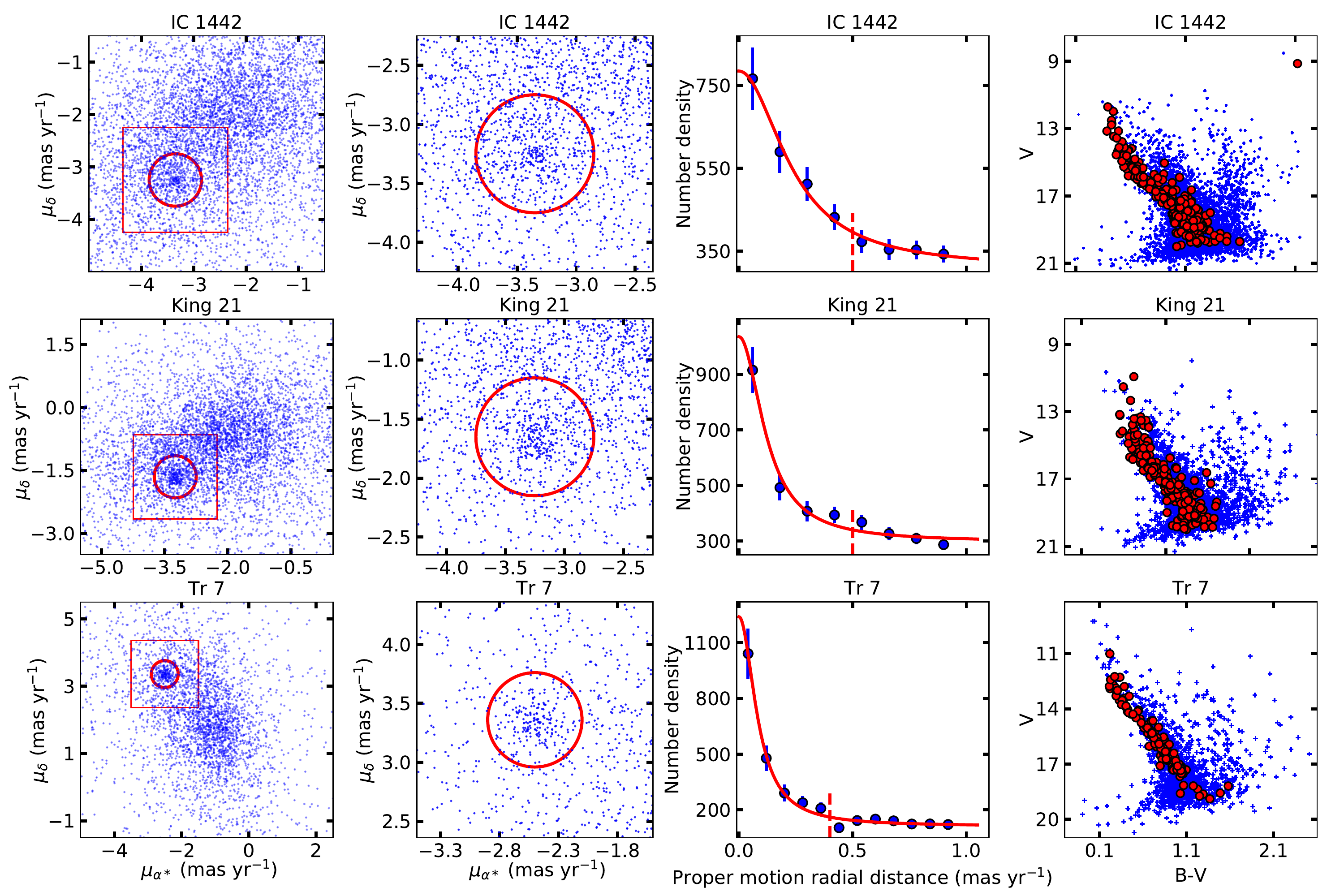}
   }
 \caption{Plots of vector-point diagram (1st column), zoomed view of rectangle in vector-point diagram (2nd column), radial stellar density distribution in proper motion plane (3rd column), and Colour-magnitude diagram (4th column) for the clusters IC 1442, King 21 and Tr 7. The probable members are encircled by red circle in the vector-point diagrams. The member stars are shown by red points and the field stars are denoted by blue '+' sign in the colour-magnitude diagrams.}
 \label{vpd_rdp}
\end{figure*}
The main obstacle in the study of open clusters is field stars contamination. It is necessary to separate field stars from the cluster member stars to estimate the physical and dynamical parameters of the clusters \citep[e.g.,][]{2008MNRAS.386.1625C,2018MNRAS.478.5184D,2018MNRAS.481.3887D}. The member stars in the open clusters are more concentrated in the proper motion plane than the field stars therefore method based on proper motions are useful in identifying members in a cluster \citep{2019MNRAS.488.1635A,2020MNRAS.493.3473A}. The vector-point diagram (VPD) based on proper motions effectively separate the member stars from the field stars in a cluster region \citep{2018MNRAS.481.3887D,2019A&A...623A..22S}. The \textit{Gaia} DR2 catalog provides kinematic data with unprecedented astrometric precision due to which membership determination based on kinematic data of the clusters has become more reliable \citep{2018A&A...618A..59C, 2019A&A...627A..35C, 2019ApJS..245...32L}. Therefore, we used data from the \textit{Gaia} DR2 archive in the kinematic study, membership determination, and distance calculation for the clusters. The VPDs  based on proper motions in the RA-DEC plane for the three selected clusters are shown in Figure~\ref{vpd_rdp}. As it can be seen from Figure~\ref{vpd_rdp} that cluster stars encircled by red circle are separated from field stars. The centers of the circles are determined by maximum density method in proper motion plane \citep{2020MNRAS.492.3602J}. The values of the centers are found to be at ($\mu_{\alpha*}$, $\mu_{\delta}$) = (-3.35 mas yr$^{-1}$, -3.25 mas yr$^{-1}$), (-3.25 mas yr$^{-1}$, -1.65 mas yr$^{-1}$), and (-2.50 mas yr$^{-1}$, 3.36 mas yr$^{-1}$) for the clusters IC 1442, King 21, and Tr 7, respectively. The radii of the circles are estimated from the plots of stellar density as a function of radial distance from the centers in the proper motion plane. The stellar density profile fitted with a function similar to the radial density profiles of open clusters is shown in Figure~\ref{vpd_rdp}. We took radius as the distance from the center where cluster star density falls close to the field star density in proper motion plane. The derived radii of the circles in the VPDs are 0.5, 0.5, and 0.4 mas yr$^{-1}$ for the clusters IC 1442, King 21, and Tr 7, respectively. We called stars within the red circles as probable member stars. We found a total of 486, 454, and 225 probable member stars for the clusters IC 1442, King 21, and Tr 7, respectively. To avoid any field star contamination and quantify membership of stars we calculated membership probabilities of stars in the cluster regions. We estimated membership probabilities using a statistical approach based on PMs of stars as also described in previous studies \citep{1971A&A....14..226S,2019MNRAS.487.3505M,2020MNRAS.494.4713M}. 

\begin{figure}
   \centering
  \includegraphics[width=8.0 cm,height=9.0 cm]{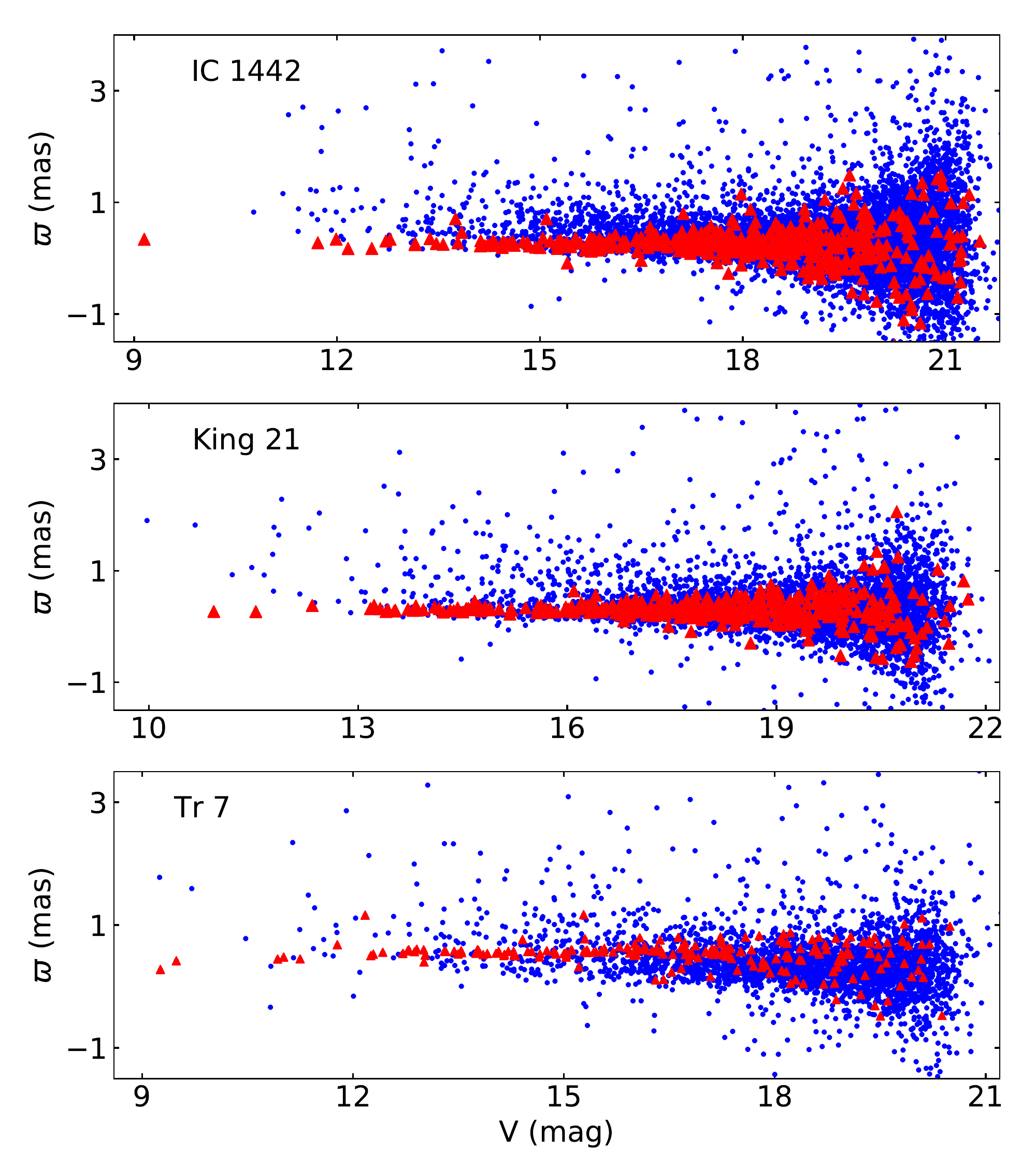}
 \caption{Parallax versus V magnitude plots for the clusters IC 1442, King 21, and Tr 7. The probable members and field stars are denoted by red triangles and blue circular points, respectively.}
 \label{V_paralx}
\end{figure}

\begin{figure}
 \hbox{
  \includegraphics[width=8.5 cm,height=5.0 cm]{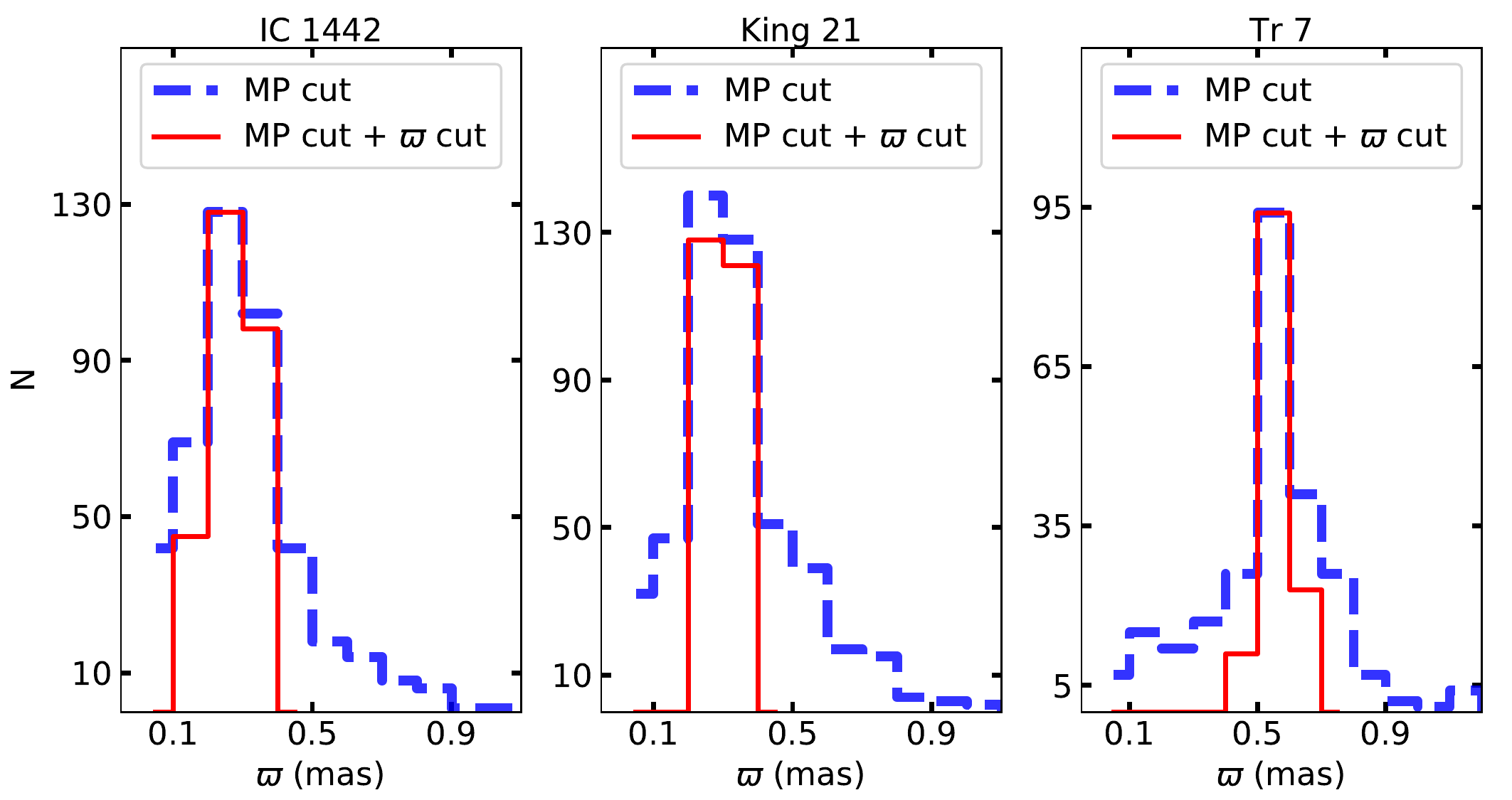}
 }
 \caption{Plots of histograms of parallax for IC 1442, King 21, and Tr 7. The histograms for stars having membership probability (MP) above 60$\%$ are shown by blue dashed line while histograms for stars satisfying both the MP cut and the parallax cut are denoted by red solid line.}
 \label{pm_cuts}
\end{figure}

The member stars of a cluster are gravitationally bound so they show very narrow spread in parallaxes than field stars in parallax ($\varpi$) versus magnitude plots \citep{2020MNRAS.493.3473A}. Therefore we applied parallax cut-off on the stars after taking membership probability cut-off to be 60$\%$. The plots for parallaxes as function of magnitude for the clusters are shown in Figure~\ref{V_paralx}. We identified stars having membership probability more than 60$\%$ and parallax within 1$\sigma$ of mean parallax obtained for the cluster region stars of the VPD as member stars for the clusters IC 1442 and King 21. However, in the case of Tr 7, the stars having membership probability more than 60$\%$ and parallax within 2$\sigma$ of the mean parallax were identified as member stars. The relaxed cut-off in parallax for the cluster Tr 7 is due to the fact that the cluster have relatively larger average parallax value.  We found 271, 249, and 128 stars satisfying membership probability and parallax cut-off in the clusters IC 1442, King21, and Tr 7, respectively. We plotted CMD from these stars. We removed stars which were outlier on the CMD and also have radial distance from the center greater than 0.3 mas yr$^{-1}$ in the proper motion plane. We also removed few CMD outlier stars as they were blended by neighbouring stars. Finally, We identified a total of 263, 244, and 128 stars to be member stars in the clusters IC 1442, King 21 and Tr 7, respectively. \citet{2020A&A...633A..99C} identified 116 and 118 member stars in the clusters King 21 and Tr 7, respectively. In Figure~\ref{pm_cuts}, we have shown histograms of parallax of stars satisfying the membership probability cut-off as well as member stars satisfying both the probability and the parallax cut-offs. One can see that parallax cut-off further refines our membership selection. The average value of PMs calculated using these member stars are given in Table~\ref{pm}. The estimated values of PMs are in agreement with the values (-3.259 mas yr$^{-1}$, -1.675 mas yr$^{-1}$) and (-2.484 mas yr$^{-1}$, 3.329 mas yr$^{-1}$) for King 21 and Tr 7, respectively reported by \citet{2020A&A...633A..99C}. The mean proper motion values for the member stars as well as field stars obtained in the present study are given in Table~\ref{pm}. Only those stars who have been identified as members are used in the following analysis.  

\input{table03}

\section{CLUSTER PARAMETERS}\label{param}

\subsection{Reddening law}\label{rv}

\begin{figure}
\includegraphics[width=9.0 cm, height=6.0 cm]{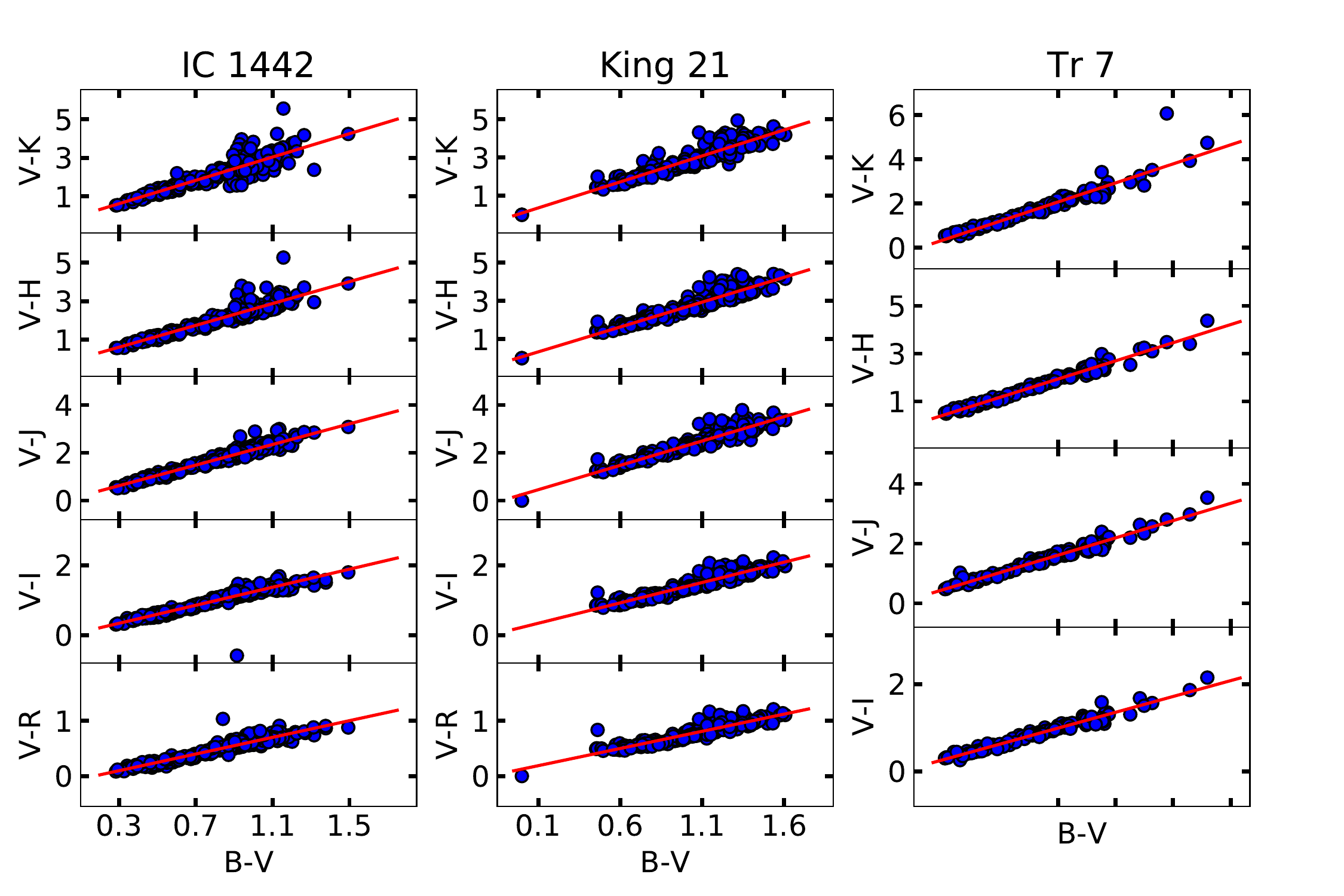}
\vspace{-0.3cm}
\caption{The (V-$\lambda$)/(B-V) diagrams for member stars in the clusters IC 1442, King 21, and Tr 7. The best fit for the slopes are presented by red solid lines.}
\label{tcd}
\end{figure}

The dust and clouds present in the line of sight cause extinction of the light reaching us by absorbing and scattering the light. It is found from many previous studies that the normal reddening law of R$_{v}$=3.1 is not always valid  \citep{2016ApJ...818...59D,2019A&A...623A.159P}. The lines of sight that pass through dust and dense clouds may have different R$_{v}$ values specially in case of young open clusters \citep{1978ApJ...223..168S,2012AJ....143...41H,2015MNRAS.446.3797H}. Therefore, the estimation of R$_{v}$ values for the young clusters which are likely to be associated with gas and dust is necessary in calculation of interstellar extinction in the direction of clusters. The slope of the (V-$\lambda$)/(B-V) two-colour diagrams (TCDs) are used to find extinction due to interstellar gas and dust \citep{1990A&A...227..213C,2019A&A...623A.159P}. The plots of TCDs (V-R), (V-I), (V-J), (V-H), and (V-K) versus (B-V) for the member stars are shown in Figure~\ref{tcd}. The value of total-to-selective extinction, R$_{cluster}$, in direction of the cluster was calculated using following approximate relation given by \citet{1981A&AS...45..451N}:

 \begin{center}
 R$_{cluster}$ = $\frac{m_{cluster}}{m_{normal}} \times R_{normal}$
\end{center}       
 where, R$_{normal}$ is normal value of total-to-selective extinction taken to be 3.1 \citep{1989ApJ...345..245C}. m$_{normal}$ and m$_{cluster}$ are the slopes of linear fit to TCDs for the normal extinction and  the extinction in the direction of the clusters, respectively. The values of m$_{normal}$ and m$_{cluster}$ are given in Table~\ref{rv_slope}. The mean values of R$_{cluster}$ are calculated to be 3.8, 3.4, and 3.3 in direction of IC 1442, King 21, and Tr 7, respectively. Thus, we found anomalous reddening law for the clusters and the higher values of R$_{v}$ indicate presence of larger dust grain size than the size in diffuse interstellar medium in the direction of these clusters \citep{2016ApJ...818...59D,2019A&A...623A.159P}. The presence of the larger dust grain size may be due to destruction of the small grains by radiation emitted from O-B type stars or growth of the grain \citep{2016ApJ...818...59D,2019A&A...623A.159P}.

\input{table04}

\subsection{Reddening determination}
\subsubsection{Reddening in optical bands}\label{ebv}

\begin{figure}
\includegraphics[width=7.5cm, height=9.0cm]{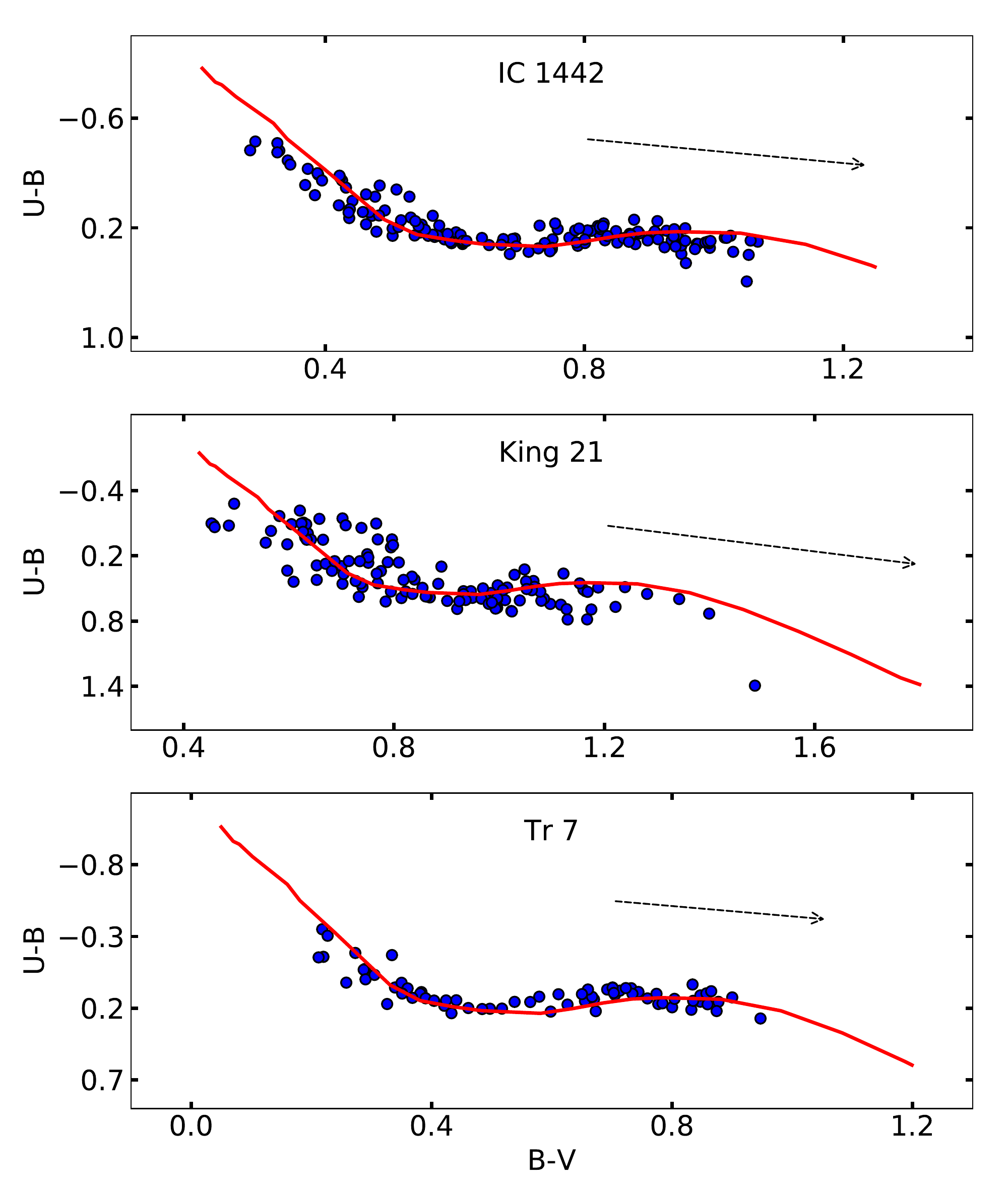}
\caption{The (U-B) versus (B-V) diagrams for IC 1442, King 21, and Tr 7. The best fit ZAMS for isochrone given by \citet{Schmidt-Kaler} are shown by solid red lines. The dashed arrows present the reddening vector.}
\label{reddening}
\end{figure}

The physical parameters of the clusters like age and distance are derived through fitting theoretical isochrones on the CMD constructed from member stars. A prior estimation of reddening is necessary in order to accurately determine the age and distance of the selected clusters from isochrone fitting on CMD. The reddening, E(B-V), can be determined by fitting intrinsic zero-age main sequence (ZAMS) isochrone on the (U-B)/(B-V) TCD \citep{1994ApJS...90...31P}. We fitted intrinsic ZAMS isochrones given by \citet{Schmidt-Kaler} to the (U-B)/(B-V) TCDs of the clusters IC 1442, King 21, and Tr 7 as shown in Figure~\ref{reddening}. The best fit of the intrinsic ZAMS isochrone was achieved by shifting E(B-V) along different values of the reddening vector $\dfrac{E(B-V)}{E(U-B)}$.  The reddening vector was found to be 0.44$\pm$0.03, 0.60$\pm$0.04, and 0.36$\pm$0.03 for the clusters IC 1442, King 21, and Tr 7, respectively. The reddening vector for these clusters are shown by black dashed arrow in Figure~\ref{reddening}. We estimated E(B-V) to be 0.54$\pm$0.04, 0.76$\pm$0.06, and 0.38$\pm$0.04 mag for IC 1442, King 21, and Tr 7, respectively. Our estimated value of the reddening for King 21 is less than 0.89 mag and 0.85$\pm$0.05 mag reported by \citet{1984BASI...12..217M} and \citet{2007A&A...462..591N}, respectively. The mean reddening, E(B-V), for the cluster Tr 7 was derived as 0.29 mag by \citet{1972A&AS....7..133V} which is lower than the value 0.38 mag estimated by us.

 \subsubsection{Reddening in near-IR bands}\label{ejhk}
 
\begin{figure}
\includegraphics[width=8.5cm, height=4.0cm]{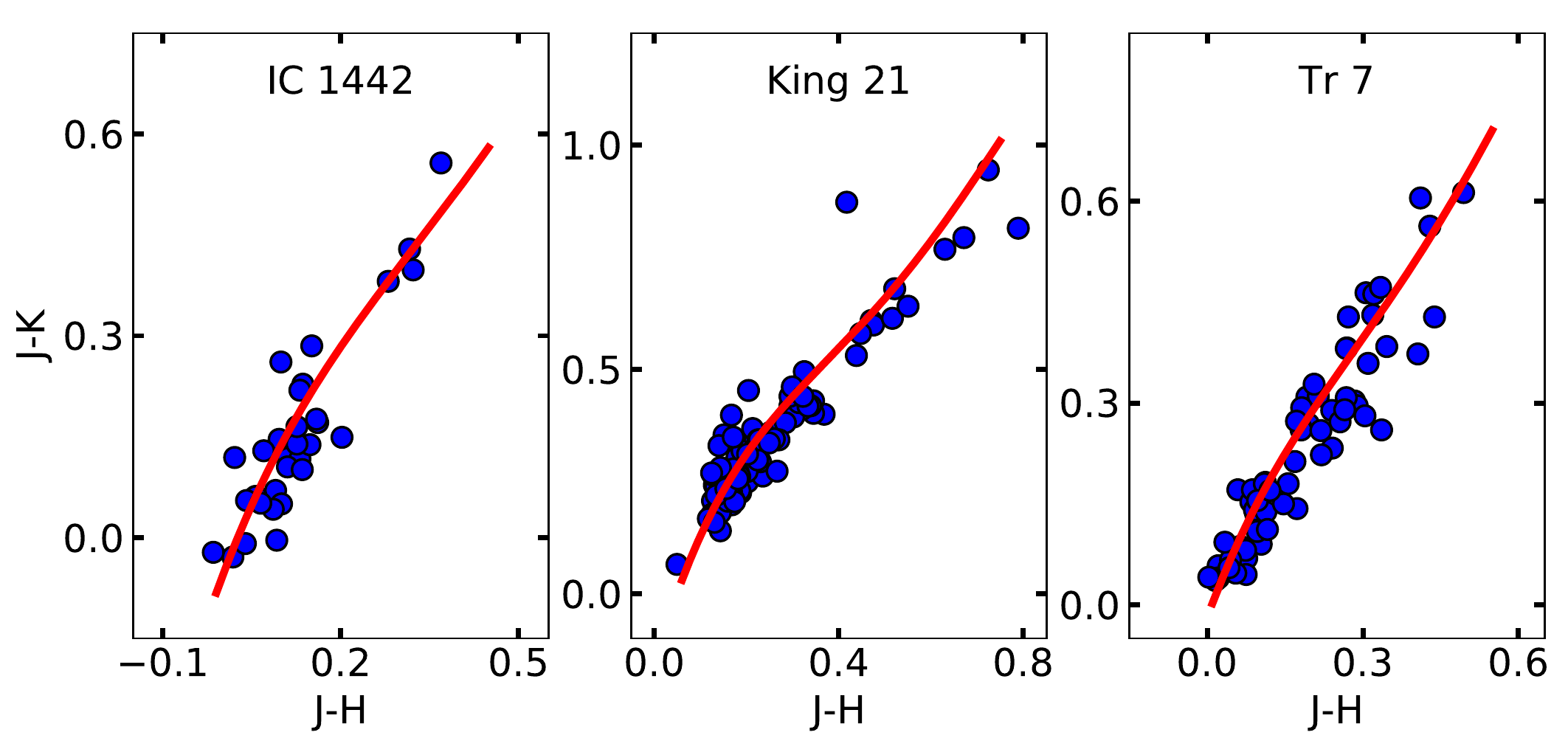}
\caption{The (J-K) versus (J-H) diagrams for IC 1442, King 21, and Tr 7. The best fit isochrone given by \citet{2017ApJ...835...77M} are shown by solid red lines.}
\label{jh_jk}
\end{figure}
The near-IR photometry has been useful in the study of interstellar extinction for open clusters \citep{2011RMxAA..47..127T}. The reddening E(B-V) can also be obtained using near-IR TCDs \citep{1989ApJ...345..245C,2005AJ....129..656C,2011RMxAA..47..127T}. We derived E(B-V) values from the near-IR (J-H)/(J-K) two-colour diagrams using the following equations:
$$E(J-H) = 0.273 \times E(B-V)$$
The (J-H)/(J-K) two-colour diagrams fitted with isochrone provided by \citet{2017ApJ...835...77M} are shown in Figure~\ref{jh_jk}. We used only those member stars which have photometric uncertainty less than 0.07 mag in J, H, and K bands to avoid uncertainty due to high error in magnitudes. We found the colour ratio $\frac{E(J-H)}{E(J-K)}$ to be 0.65, 0.62, and 0.60 for IC 1442, King 21, and Tr 7, respectively. The colour ratio, $\frac{E(J-H)}{E(J-K)}$, found by us are greater than the normal interstellar extinction ratio of 0.55 given by \citet{1989ApJ...345..245C}. We achieved best fit values of E(J-H) to be 0.15, 0.20, and 0.10 mag which corresponds to E(B-V) values of 0.55, 0.73, and 0.37 mag for IC 1442, King 21 and Tr 7, respectively. The E(B-V) values derived from the near-IR TCDs reconfirms our finding of E(B-V) values 0.54$\pm$0.04, 0.76$\pm$0.06, and 0.38$\pm$0.04 mag estimated from the optical (U-B)/(B-V) TCDs for IC 1442, King 21, and Tr 7, respectively.

\subsection{Age and Distance determination}
\subsubsection{Distance through parallax}\label{Dparalx}
\begin{figure}
\includegraphics[width=8.5cm, height=4.0 cm]{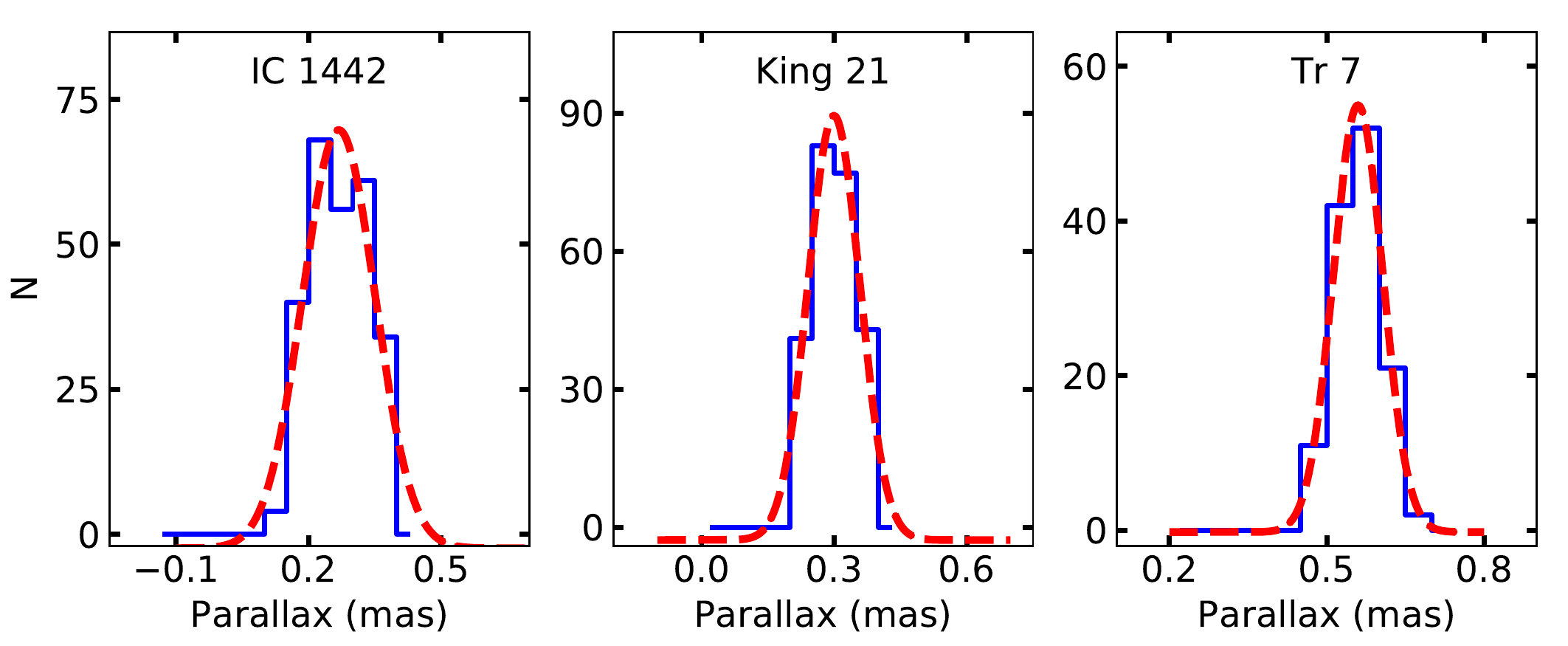}
\caption{The plots for histograms of parallax with bin size 0.05 mas for IC 1442, King 21, and Tr 7. The Gaussian fitted on the histograms are shown by red dashed curves.}
\label{paralx}
\end{figure}

The kinematic data from \textit{Gaia} DR2 has been used in various useful studies like membership determination, search of new open clusters, determination of distances \citep{2020A&A...633A.146A,2020arXiv200107122C}. We used parallax from \textit{Gaia} DR2 data to estimate distances to the open clusters. The presence of systematic offsets in parallaxes of \textit{Gaia} DR2 has been reported in previous studies \citep{2018A&A...616A...2L, 2018A&A...616A...9L,2018ApJ...862...61S}. It is preferred to estimate average distance to an open cluster from the average value of parallax of member stars rather than averaging distance of individual stars inferred from their parallaxes \citep{2018A&A...616A...9L}. We estimated the mean value of the parallax angle for the clusters by fitting Gaussian on the histograms. The plots for the histograms with bin size of 0.05 mas and Gaussian fit are shown in Figure~\ref{paralx}. In this way we estimated mean parallax angle as 0.268$\pm$0.082, 0.299$\pm$0.058, and 0.560$\pm$0.047 mas. The calculated values of parallax angles are in agreement with the parallax 0.305 and 0.554 obtained by \citet{2020A&A...633A..99C} for the clusters King 21 and Tr 7, respectively. We calculated distances by inverting the mean parallaxes of clusters after applying offset of -0.082 mas in the estimated mean parallaxes \citep{2018ApJ...862...61S} and obtained a distance of 2857$\pm$708, 2625$\pm$409, and 1558$\pm$115 pc for IC 1442, King 21, and Tr 7, respectively. A direct estimation of distance from \textit{Gaia} DR2 parallaxes using parallax-inversion method may results in a biased distance estimate \citep{2018A&A...616A...9L}. \citet{2015PASP..127..994B} suggested a method based on  probabilistic analysis to calculate more accurate distance than distance estimated by inverting parallax. We found distances after applying method as described in \citet{2018AJ....156...58B} as 2847$\pm$238, 2622$\pm$156, and 1561$\pm$74 pc for the clusters IC 1442, King 21, and Tr 7, respectively. There is not any significant discrepancy in distances calculated using these two methods in the present study. However, our estimated distances are lower than the distances $\sim$2994 pc and $\sim$1714 pc obtained by \citet{2020A&A...633A..99C} for the clusters King 21 and Tr 7, respectively. \citet{1970A&A.....8..213Y} reported much lower distance  of 1810 pc for IC 1442. The distance of King 21 is estimated to be 3.41 kpc by \citet{2008MNRAS.388.1879M} higher than our estimated value. The distance to cluster Tr 7 is found to be 1.60 kpc by \citet{2005ApJS..161..118M} which is consistent with our estimate.

\begin{figure*}
\includegraphics[width=16.0cm, height=5.5cm]{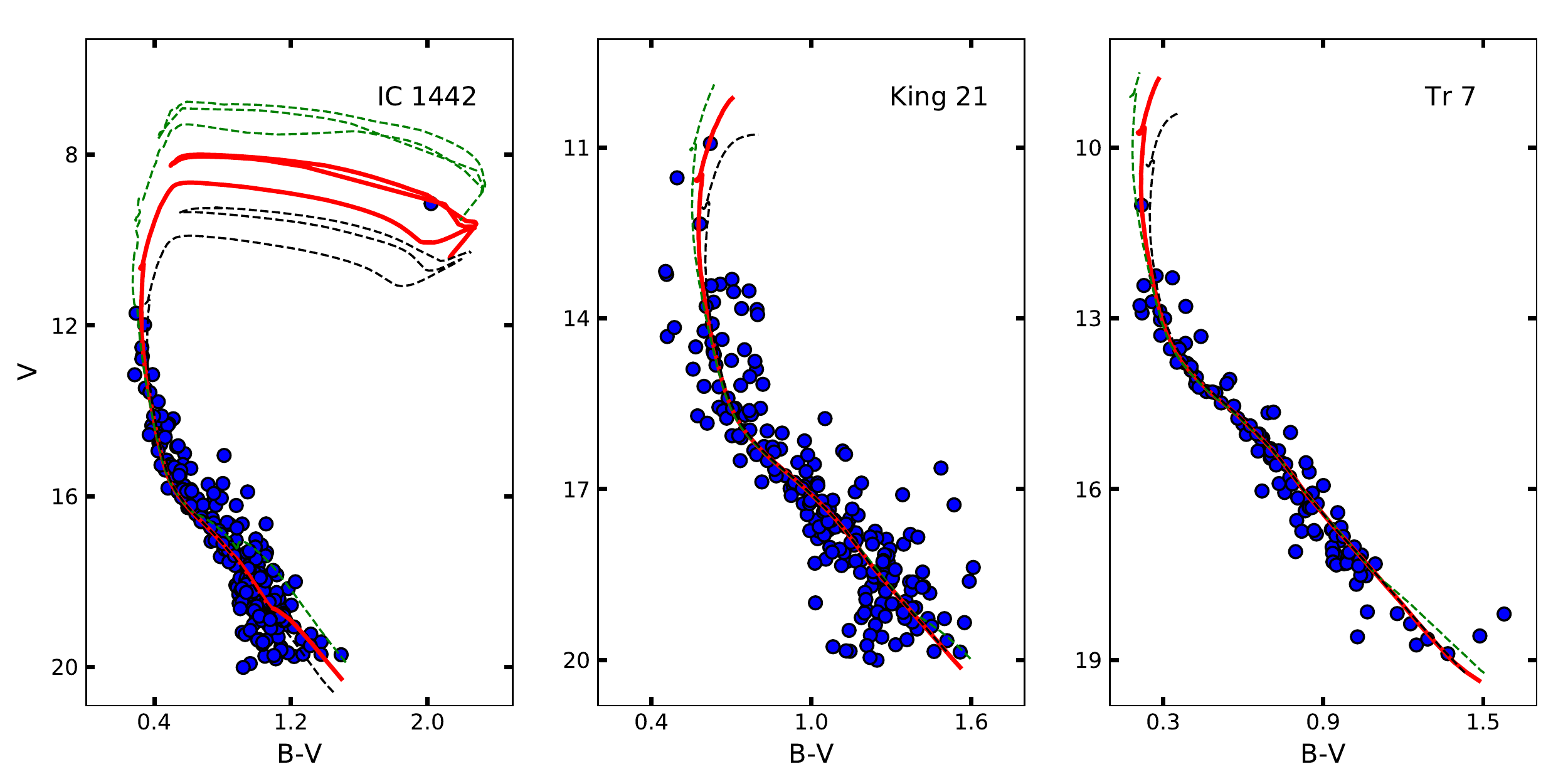}
\caption{The (B-V)/V CMDs for IC 1442, King 21, and Tr 7. The best fit \citep{2017ApJ...835...77M} isochrone are presented by red solid curves. The black and the green dashed lines present \citep{2017ApJ...835...77M} isochrones corresponding to the upper and lower age limits of the clusters.}
\label{bv_v}
\end{figure*}
\subsubsection{Age through CMD}

The CMDs of the clusters have been very useful tool in identification of the member stars and determination of the cluster parameters. These are also used in the study of blue stragglers, rotation of stars, and multi-population in the clusters \citep{2018ApJ...869..139C,2019MNRAS.490.2414P,2020AJ....159...59R}. In this study we used (B-V)/V CMDs to derive age of the clusters. The fitting of isochrone on the CMD is affected by metallicity, reddening, age corresponding to isochrone, and apparent distance modulus. We fitted theoretical isochrones of solar metallicity given by \citet{2017ApJ...835...77M} on CMDs of the clusters IC 1442, King 21, and Tr 7.  The reddening values derived using (U-B)/(B-V) were used for the isochrone fitting. We calculated apparent distance modulus using distance and total-to-selective extinction estimated in Section~\ref{Dparalx} and ~\ref{rv}. We found apparent distance modulus to be 14.32, 14.68, and 12.22 mag for IC 1442, King 21, and Tr 7, respectively. We fitted isochrones of different ages on the CMDs and the plots of best fit CMDs are shown in Figure~\ref{bv_v} for these clusters. There is broadening in the CMDs for fainter stars which can be due to larger photometric errors in the magnitudes towards $\sim$20 mag stars. Through isochrone fitting to the CMDs we derived log(Age) = 7.40$\pm$0.30, 7.70$\pm$0.20, and 7.85$\pm$0.25 years for the clusters IC 1442, King 21, and Tr 7, respectively. The age of 50 Myr found by us is close to 30 Myr estimated by \citet{2008MNRAS.388.1879M} for King 21. However, \citet{2007A&A...462..591N} found lower age of $\sim$16 Myr for King 21. Our derived value of log(Age) is greater than the value 7.40 yr estimated by \citet{2005ApJS..161..118M} for the cluster Tr 7. The structural and physical parameters estimated in the present study are summarized in Table~\ref{clust_par}.  

\input{table05}
\section{Dynamical study of the clusters}\label{dynamic}
 
\subsection{Luminosity function}

\input{table06}

The luminosity function (LF) is defined as number of stars belonging to each magnitude bins. The derivation of the LF is prone to uncertainty caused by data incompleteness specially for faint stars. To avoid inaccuracy in LF we estimated completeness of our photometric data as discussed in Sect.~\ref{complete}.  We calculated LF from member stars detected in V band by dividing number of stars detected in each magnitude bin with CF for the given bin. The completeness of our data is above 90$\%$ upto 18, 19, and 19 mag in V band for the clusters IC 1442, King 21, and Tr 7, respectively. The derived LF for these clusters is given in Table~\ref{lf}.

\begin{figure*}
\centering
\includegraphics[width=16.0 cm, height=4.0 cm]{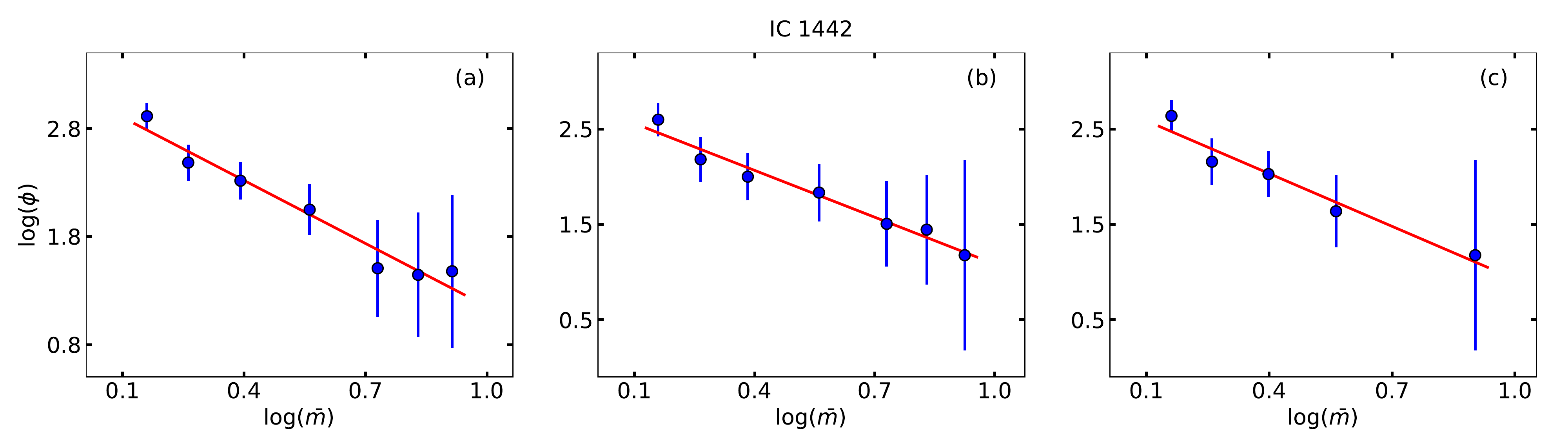}
\includegraphics[width=16.0 cm, height=4.0 cm]{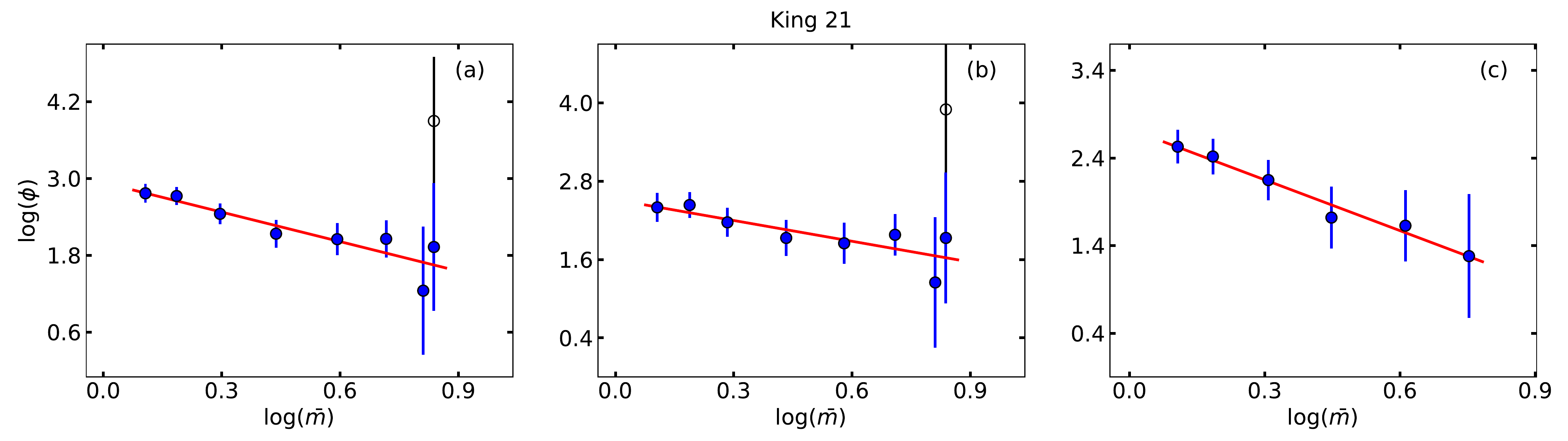}
\includegraphics[width=16.0 cm, height=4.0 cm]{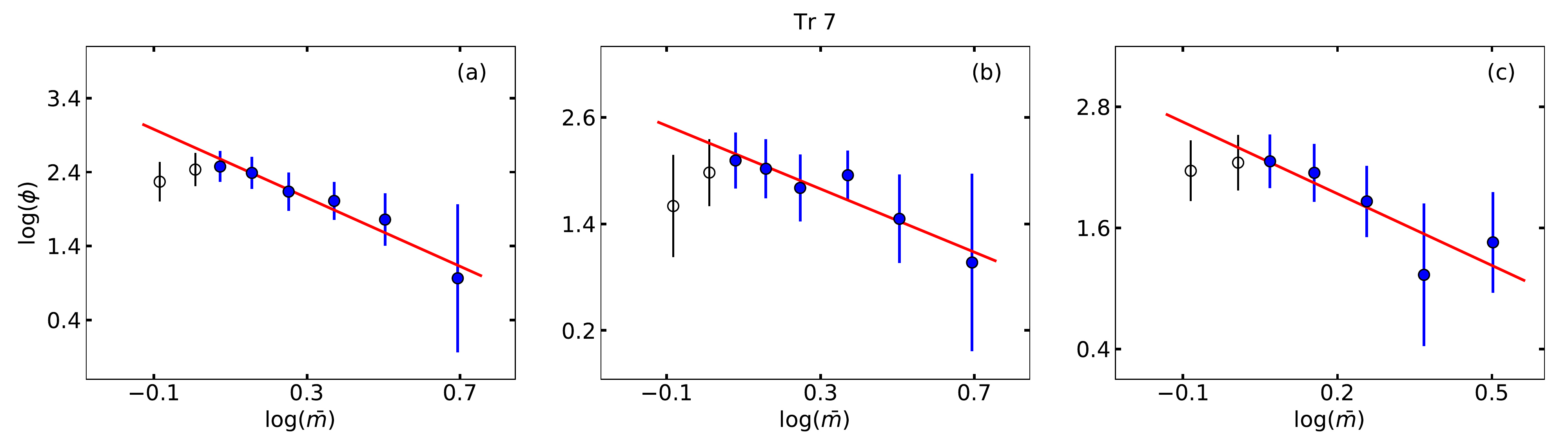}
\caption{The plots of MF slopes for IC 1442 (upper panel), King 21 (middle panel), and Tr 7 (lower panel) in (a) entire, (b) inner, and (c) outer region. The points presented by black open circles are excluded from the estimation of MF slopes.}
\label{mf}
\end{figure*}

\subsection{Mass function}

\input{table07}

The initial mass function (IMF) help us to have insight of initial condition of star formation events. The young open clusters are essential in the study of IMF as their present-day mass function are nearly associated with IMF \citep{2018A&A...614A..43D}. The value of the MF slope of a cluster may be affected due to the dynamical evolution and mass segregation \citep{2019MNRAS.489.2377S}. The MF is defined as the relative number of stars per unit mass and often expressed by power law N (log m) $\propto$ m$^{\Gamma}$. The slope of MF is denoted by $\Gamma$ and calculated using formula: ${\Gamma}$ = $d \ log \ N(log \ m)/d \ log \ m$, where N log(m) denotes the number of stars per unit logarithmic mass. We calculated mass function from the luminosity function. The mass of the stars were derived by fitting isochrones of \citet{2017ApJ...835...77M} on the CMD of the clusters using reddening, age, and apparent distance modulus derived in Sect.~\ref{param}. We used linear least square fit to derive the MF slopes. We first estimated the MF slopes in the entire region observed for the respective clusters. We then derived MF slopes for the inner and the outer regions of the clusters separately to study spatial variation in the MF slopes within the cluster. We took a circular region as inner region around the cluster centers of the radii equal to 6$^{\prime}$.5, 4$^{\prime}$.0, and 4$^{\prime}$.0 for the clusters IC 1442, King 21, and Tr 7, respectively. The outer region is defined as the observed entire region minus the inner region. We found MF slopes in the entire region of clusters to be -1.94$\pm$0.18 in mass range 9.31-1.32 M$_{\odot}$, -1.61$\pm$0.35 in the range 6.89-1.16 M$_{\odot}$, and -2.28$\pm$0.26 in the range 4.96-1.07 M$_{\odot}$ for the clusters IC 1442, King 21, and Tr 7, respectively. The plots for linear least square fit of MF slopes are shown in Figure~\ref{mf}. We excluded two very off points from the MF slopes fitting for Tr 7 as shown with black open circles in Figure~\ref{mf}. These two points corresponds to mass ranges 1.07-0.90 M$_{\odot}$ and 0.90-0.76 M$_{\odot}$ which are exactly below turn-over mass $\sim$1 M$_{\odot}$ from where MF slope have been reported to follow different slope than high mass stars \citep{2001MNRAS.322..231K,2005A&A...437..483B,2018A&A...620A..39J,2019MNRAS.489.2377S,2020MNRAS.494.4713M}. However, \citet{2016EAS....80...73M} report turn-over mass to be 0.5 M$_{\odot}$ and \citet{2013MNRAS.434.3236K} report 0.65 M$_{\odot}$ to be the turn-over mass. The two-step power law nature of MF slope may be reason behind the different trends for these two points but our data is not complete beyond mass range 0.90-0.76 M$_{\odot}$ in order to analyse the MF slope nature for lower mass ranges. The derived slopes and corresponding uncertainty of linear regression solution for the entire, inner, and outer regions of the clusters are given in Table~\ref{mf_slope}. We also found MF slopes  in the outer regions of the selected clusters to be relatively steeper than MF slopes in the inner regions which indicates relatively high concentration of high-mass stars in the inner region due to mass segregation in the cluster \citep{2013MNRAS.434.3236K,2020MNRAS.494.4713M}. We further investigated the mass segregation through statistical approach in the next section.

\subsection{Mass segregation}\label{segre}

\begin{figure}
\centering
\includegraphics[width=8 cm, height=12 cm]{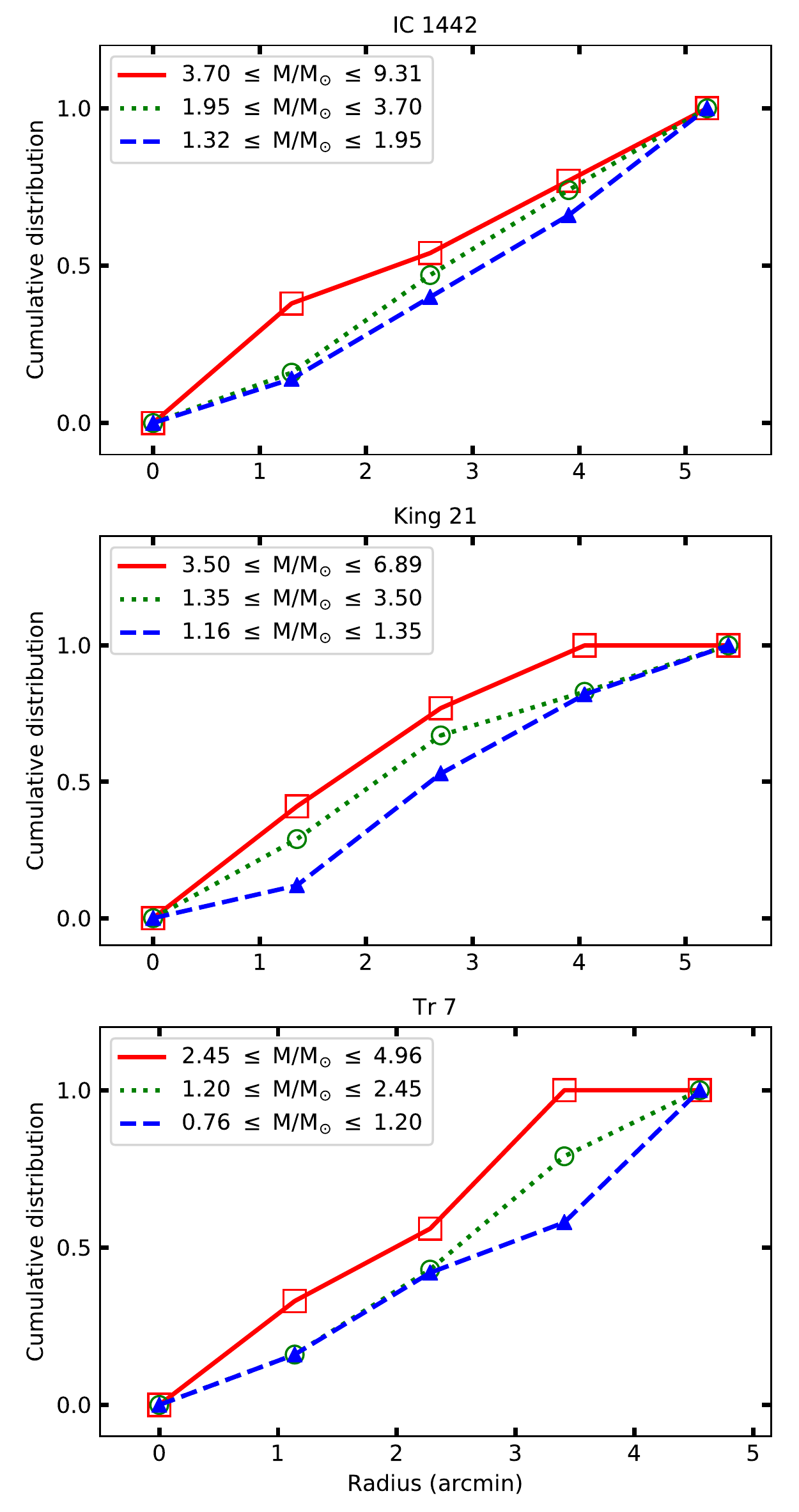}
\caption{The figure presents variation of cumulative distribution of number of stars with radial distance in IC 1442, King 21, and Tr 7, respectively. The red solid lines, green dotted lines, and blue dashed lines correspond to the high, intermediate, and low mass ranges, respectively. The mass ranges are mentioned in upper left of the each plots.}
\label{cum_dist}
\end{figure}

Open clusters are known to have mass segregation \citep{2018MNRAS.473..849D,2019MNRAS.490.2521H,2019A&A...629A.135D}. The mass segregation is the phenomenon in which bright and massive stars are more centrally concentrated than faint and low mass stars in clusters. It is still debated whether mass segregation takes place due to the dynamical evolution of the cluster or star formation process itself or a combination of both \citep{2018MNRAS.473..849D,2018A&A...615A...9P}. In dynamical evolution of clusters  massive stars transfer their kinetic energy to low mass stars during equipartition of energy and accumulate in the central region of the cluster while low-mass stars drift towards outer region \citep[e.g.,][]{2009MNRAS.395.1449A}. The mass segregation caused by star formation process itself takes place due to formation of massive stars preferentially in the central region of the cluster \citep{2007MNRAS.381L..40D,2010MNRAS.405..401D}. The young open clusters are of importance in studies of mass segregation as they are nearly associated with formation process and early dynamical evolution \citep{2019A&A...626A..79P,2018A&A...615A...9P}. The cumulative distribution of stars with radial distance is used to study the mass segregation in star clusters. We derived the cumulative radial distribution of member stars in different mass ranges to study the possible mass segregation in the clusters as shown in Figure~\ref{cum_dist}. We also corrected our data for incompleteness in the calculation of cumulative distributions. We took high, intermediate, and low mass ranges which are also mentioned in Figure~\ref{cum_dist}. The mass ranges were taken as to have statistically enough stars in each bins. As it can be noticed from the figure that the massive stars are dominantly distributed in the central part while low mass stars are distributed in the outer region of the clusters. We also performed Kolmogorov-Smirnov (K-S) test on the these mass range to examine whether they are statistically different samples or not and we conclude with above 99$\%$ confidence level that mass segregation effect is present in the clusters IC 1442, King 21, and Tr 7. This finding of mass segregation through radial cumulative method supports our conclusion that steepening of the MF slopes in outer region indicates mass segregation present in the clusters.

\subsection{The dynamical relaxation time}

The dynamical relaxation time is calculated to find whether dynamical relaxation or star formation itself causes mass segregation in the clusters \citep{2007ApJ...655L..45M,2019MNRAS.488.1635A}. The dynamical relaxation time, T$_{E}$, is the time required in which cluster members exchange energies and the velocity distribution of stars  in the cluster approaches to Maxwellian equilibrium velocity distribution. T$_{E}$ is calculated using following formula given by \citet{1971ApJ...164..399S}:
$$
T_E = \frac{8.9 \times 10^5 (N R_h^3/\bar{m})^{1/2}} {\log(0.4N)}
$$
where N is the total number of member stars and R$_{h}$ is half-mass radius (in parsecs) of the cluster. The $\bar{m}$ is mean mass of the member stars in solar mass units. T$_{E}$ is derived relaxation time in years. The R$_{h}$ is calculated using the cumulative mass of the stars with increasing radial distance from the center of the cluster. The R$_{h}$ is the radial distance within which half of the total cluster mass lies. We found R$_{h}$ values to be 5$^{\prime}$.9 (4.9 pc), 3$^{\prime}$.0 (2.3 pc), and 3$^{\prime}$.7 (1.7 pc) and corresponding T$_{E}$ values as 74, 26, and 34 Myr for the clusters IC 1442, King 21 and Tr 7, respectively. As the relaxation times for King 21 and Tr 7 are found to be smaller than their ages we can conclude that these clusters are dynamically relaxed and mass segregation in these clusters are caused by dynamical relaxation. The cluster IC 1442 has T$_{E}$ larger than its age therefore the mass segregation in it raises the speculation of segregation due to star formation process itself. However, early mass segregation in IC 1442 can be attributed to rapid dynamical evolution occurring even in very young clusters as suggested in some past studies \citep{2007ApJ...655L..45M,2009ApJ...700L..99A}.   

\subsection{Tidal radius}

The tidal interactions play important role in the dynamic evolution of open clusters \citep{2019A&A...621L...3M,2020MNRAS.492.4959J,2020MNRAS.493.3473A}. The structures of open clusters are also influenced by tidal interactions as strong tidal effect causes compact structure of clusters \citep{2020MNRAS.493.3473A}. Young open clusters orbits closer to the dense Galactic plane so they are more likely impacted by the tidal interactions. The tidal radius of a cluster is the distance from its center where the gravitational acceleration of the cluster becomes equal to the tidal acceleration of the parent Galaxy \citep{1957ApJ...125..451V}. The member stars of the clusters lie within the tidal radius and are generally gravitationally bound to the cluster. The stars which lie outside the tidal radius of the cluster are more influenced by the external potential of the Galaxy in comparison to the effective potential of the cluster. We used the following relation given by \citet{2000ApJ...545..301K} to calculate tidal radius R$_{t}$:
$$
R_{t} = \left(\frac{M_{C}}{2 M_{G}}\right)^{1/3}\times R_{G}
$$
Where M$_{C}$ and R$_{G}$ are the total mass and the Galactocentric distance of the clusters. The M$_{G}$ is the Galactic mass contained within R$_{G}$. We obtained the total masses for the clusters as 483.16, 462.56, and 186.00 M$_{\odot}$ for the clusters IC 1442, King 21, and Tr 7, respectively by fitting \citet{2017ApJ...835...77M} isochrone on the CMD of cluster employing the member stars only. We used the distance between the Galactic center and the sun as R$_{0}$=8.0 kpc \citep{2017Ap&SS.362...79V,2018PASP..130b4101C}. The estimated values of R$_{G}$ are found to be 9.31, 9.44, and 9.34 kpc for the clusters IC 1442, King 21, and Tr 7, respectively.

The Galactic mass contained within R$_{G}$ is calculated using the following relation given by \citet{1987ARA&A..25..377G}:
$$
M_{G} = 2 \times 10^{8} M_{\odot} \left(\frac{R_{G}}{30pc }\right)^{1.2}
$$
We obtained the values of M$_{G}$ calculated using above equation as $\sim$ 1.96$\times$10$^{11}$, 1.99$\times$10$^{11}$, 1.96$\times$10$^{11}$ M$_{\odot}$ for IC 1442, King 21, and Tr 7, respectively and corresponding tidal radius are calculated to be 10.70, 10.64, and 7.82 pc, respectively. The tidal radii are larger than the cluster radii estimated in Sect.~\ref{RDP} and the stars beyond the tidal radius of these clusters would be considered as gravitationally unbound to the respective cluster. The cluster parameters obtained from the dynamical study of the selected open clusters in the present study are given in Table~\ref{dyna}. The number of member stars used as mentioned in Table~\ref{dyna} are those member stars which are detected in V band.

\input{table08}
%
\section{Summary}\label{summary}
Here, we present a comprehensive photometric study of three poorly studied young open star clusters IC 1442, King 21, and Tr 7 as continuation of our effort to study some of the poorly studied open clusters in the Galaxy. We used $UBVR_cI_c$ photometric data taken with 1.3-m DFOT at Nainital supplemented with astrometric data of \textit{Gaia} DR2 and the 2MASS JHK near-IR data. These clusters have been included only in a few previous studies while no previous comprehensive deep photometry was available for any of these clusters. We used kinematic data from \textit{Gaia} DR2 in order to identify member stars in the clusters to estimate the physical parameters and understand the stellar and dynamical evolution in these three clusters. We summarize the results of the present study as following:
\begin{enumerate}

\item We derived the cluster radii from their radial density profiles as 9$^{\prime}$.5$\pm$0$^{\prime}$.5, 8$^{\prime}$.7$\pm$0$^{\prime}$.5, and 7$^{\prime}$.1$\pm$0$^{\prime}$.5 for the clusters IC 1442, King 21, and Tr 7, respectively.

\item We used \textit{Gaia} DR2 proper motions and parallaxes to determine membership of the stars present in the clusters. We identified a total of 263, 244, and 128 member stars in the clusters IC 1442, King 21, and Tr 7, respectively. All the derivation of physical as well as dynamical parameters in the present study has been carried out using these identified cluster members only.

\item We obtained the mean proper motion, $\bar{\mu}$, of the clusters IC 1442, King 21, and Tr 7 using proper motions of member stars as 4.67$\pm$0.23, 3.64$\pm$0.22, and 4.17$\pm$0.15 mas yr$^{-1}$, respectively.

\item The derived reddening values to be E(B-V) = 0.54$\pm$0.04, 0.76$\pm$0.06, and 0.38$\pm$0.04 mag using (U-B)/(B-V) diagrams for the clusters IC 1442, King 21, and Tr 7, respectively.

\item We derived reddening law using (V-$\lambda$)/(B-V) TCDs and found anomalous reddening law in the direction of these clusters as R$_{cluster}$ = 3.8, 3.4, and 3.3 for IC 1442, King 21, and Tr 7, respectively. The anomalous nature of reddening law was noticed to be consistent in the optical as well as the near-IR studies.

\item We derived age of the clusters to be log(Age/yr) = 7.40$\pm$0.30, 7.70$\pm$0.20, and 7.85$\pm$0.25 for IC 1442, King 21 and Tr 7, respectively and their distances were estimated as 2847$\pm$238, 2622$\pm$156, and 1561$\pm$74 pc using the parallax of the cluster members.

\item The MF slopes in the entire observed regions of clusters IC 1442, King 21, and Tr 7 were found to be -1.94$\pm$0.18, -1.54$\pm$0.32, and -2.31$\pm$0.29, respectively.

\item  We estimated MF slopes of the clusters separately in the inner and the outer regions of the clusters IC 1442, King 21, and Tr 7 and found steeper MF slopes in the outer regions indicating possible mass segregation in these clusters. 

\item We also studied dynamical evolution in these clusters by deriving cumulative density distribution with radial distance in different mass bins and found that massive stars dominantly distributed in the inner regions of these clusters. The concentrated distribution of massive stars in the inner regions of the clusters may be due to mass segregation. 

\end{enumerate} 

\section*{Acknowledgements}
We used data from the Two Micron All Sky Survey, which is a joint project of the University of Massachusetts; the Infrared Processing and Analysis Center/California Institute of Technology, funded by the NASA. We also used data from the European Space Agency (ESA) mission Gaia (https://www.cosmos.esa.int/Gaia), processed by the Gaia Data Processing and Analysis Consortium (DPAC, https://www.cosmos.esa.int/web/GAIA/dpac/consortium). 
\bibliographystyle{mnras}

\bibliography{main}

\end{document}

%% file: table01.tex
\hspace{-1 cm}
\begin{table} \fontsize{6.8}{7.0}\selectfont
\caption{The average photometric error ($\sigma$) with number of stars given in brackets for each magnitude bin in UBVRI bands.}
\label{mag_err}
  \begin{tabular}{l l l l l l l l l l l}  
  \hline  
 \hspace{-0.2 cm} Mag.&\multicolumn{5}{c}{IC 1442} \\
    \cmidrule(lr){2-6}
\hspace{-0.2 cm} range&\hspace{0 cm} $\sigma_{U}$&\hspace{0 cm} $\sigma_{B}$&\hspace{0 cm} $\sigma_{V}$&\hspace{0 cm} $\sigma_{R}$&\hspace{0 cm} $\sigma_{I}$ \\ \hline
  \hspace{-0.2 cm} 10-11&\hspace{-0.2 cm} -&\hspace{-0.2 cm} 0.003 (1)&\hspace{-0.2 cm} 0.030 (1)&\hspace{-0.2 cm} 0.004 (6)&\hspace{-0.2 cm} 0.008 (12)\\
  \hspace{-0.2 cm} 11-12&\hspace{-0.2 cm} 0.007 (6)&\hspace{-0.2 cm} 0.003 (6)&\hspace{-0.2 cm} 0.008 (18)&\hspace{-0.2 cm} 0.013 (24)&\hspace{-0.2 cm} 0.008 (38)\\
  \hspace{-0.2 cm} 12-13&\hspace{-0.2 cm} 0.003 (20)&\hspace{-0.2 cm} 0.007 (21)&\hspace{-0.2 cm} 0.008 (23)&\hspace{-0.2 cm} 0.013 (49)&\hspace{-0.2 cm} 0.009 (79)\\
  \hspace{-0.2 cm} 13-14&\hspace{-0.2 cm} 0.005 (22)&\hspace{-0.2 cm} 0.003 (31)&\hspace{-0.2 cm} 0.004 (85)&\hspace{-0.2 cm} 0.013 (101)&\hspace{-0.2 cm} 0.008 (152)\\
  \hspace{-0.2 cm} 14-15&\hspace{-0.2 cm} 0.003 (63)&\hspace{-0.2 cm} 0.003 (90)&\hspace{-0.2 cm} 0.006 (143)&\hspace{-0.2 cm} 0.012 (207)&\hspace{-0.2 cm} 0.009 (338)\\
  \hspace{-0.2 cm} 15-16&\hspace{-0.2 cm} 0.004 (118)&\hspace{-0.2 cm} 0.004 (145)&\hspace{-0.2 cm} 0.009 (293)&\hspace{-0.2 cm} 0.013 (424)&\hspace{-0.2 cm} 0.009 (613)\\
  \hspace{-0.2 cm} 16-17&\hspace{-0.2 cm} 0.006 (206)&\hspace{-0.2 cm} 0.005 (282)&\hspace{-0.2 cm} 0.008 (528)&\hspace{-0.2 cm} 0.013 (752)&\hspace{-0.2 cm} 0.014 (1108)\\
  \hspace{-0.2 cm} 17-18&\hspace{-0.2 cm} 0.011 (381)&\hspace{-0.2 cm} 0.008 (492)&\hspace{-0.2 cm} 0.010 (915)&\hspace{-0.2 cm} 0.016 (1256)&\hspace{-0.2 cm} 0.023 (1827)\\
  \hspace{-0.2 cm} 18-19&\hspace{-0.2 cm} 0.023 (646)&\hspace{-0.2 cm} 0.015 (874)&\hspace{-0.2 cm} 0.017 (1453)&\hspace{-0.2 cm} 0.023 (1891)&\hspace{-0.2 cm} 0.034 (1491)\\
  \hspace{-0.2 cm} 19-20&\hspace{-0.2 cm} 0.062 (1141)&\hspace{-0.2 cm} 0.028 (1358)&\hspace{-0.2 cm} 0.031 (1921)&\hspace{-0.2 cm} 0.031 (1048)&\hspace{-0.2 cm} 0.072 (161)\\
  \hspace{-0.2 cm} 20-21&\hspace{-0.2 cm} 0.133 (498)&\hspace{-0.2 cm} 0.059 (1802)&\hspace{-0.2 cm} 0.051 (743)&\hspace{-0.2 cm} 0.052 (48)&\hspace{-0.2 cm} 0.184 (3)\\
  \hspace{-0.2 cm} 21-22&\hspace{-0.2 cm} 0.373 (5)&\hspace{-0.2 cm} 0.120 (1021)&\hspace{-0.2 cm} 0.106 (26)&\hspace{-0.2 cm} -&\hspace{-0.2 cm} -\\
  \hline 
  \hspace{-0.2 cm} & \multicolumn{5}{c}{King 21} \\
  \cmidrule (lr){2-6}
  \hspace{-0.2 cm} 10-11&\hspace{-0.2 cm} -&\hspace{-0.2 cm} 0.007 (1)&\hspace{-0.2 cm} 0.006 (2)&\hspace{-0.2 cm} 0.011 (5)&\hspace{-0.2 cm} 0.003 (12)\\
  \hspace{-0.2 cm} 11-12&\hspace{-0.2 cm} 0.003 (4)&\hspace{-0.2 cm} 0.004 (3)&\hspace{-0.2 cm} 0.003 (9)&\hspace{-0.2 cm} 0.004 (15)&\hspace{-0.2 cm} 0.001 (27)\\
  \hspace{-0.2 cm} 12-13&\hspace{-0.2 cm} 0.003 (8)&\hspace{-0.2 cm} 0.002 (9)&\hspace{-0.2 cm} 0.002 (12)&\hspace{-0.2 cm} 0.002 (36)&\hspace{-0.2 cm} 0.002 (62)\\
  \hspace{-0.2 cm} 13-14&\hspace{-0.2 cm} 0.002 (11)&\hspace{-0.2 cm} 0.002 (14)&\hspace{-0.2 cm} 0.002 (47)&\hspace{-0.2 cm} 0.002 (77)&\hspace{-0.2 cm} 0.002 (129)\\
  \hspace{-0.2 cm} 14-15&\hspace{-0.2 cm} 0.003 (35)&\hspace{-0.2 cm} 0.002 (49)&\hspace{-0.2 cm} 0.003 (96)&\hspace{-0.2 cm} 0.002 (155)&\hspace{-0.2 cm} 0.003 (222)\\
  \hspace{-0.2 cm} 15-16&\hspace{-0.2 cm} 0.005 (59)&\hspace{-0.2 cm} 0.003 (93)&\hspace{-0.2 cm} 0.004 (155)&\hspace{-0.2 cm} 0.003 (267)&\hspace{-0.2 cm} 0.004 (503)\\
  \hspace{-0.2 cm} 16-17&\hspace{-0.2 cm} 0.007 (94)&\hspace{-0.2 cm} 0.004 (135)&\hspace{-0.2 cm} 0.005 (306)&\hspace{-0.2 cm} 0.005 (546)&\hspace{-0.2 cm} 0.007 (859)\\
  \hspace{-0.2 cm} 17-18&\hspace{-0.2 cm} 0.012 (148)&\hspace{-0.2 cm} 0.007 (254)&\hspace{-0.2 cm} 0.008 (522)&\hspace{-0.2 cm} 0.009 (890)&\hspace{-0.2 cm} 0.013 (1136)\\
  \hspace{-0.2 cm} 18-19&\hspace{-0.2 cm} 0.025 (292)&\hspace{-0.2 cm} 0.013 (434)&\hspace{-0.2 cm} 0.014 (851)&\hspace{-0.2 cm} 0.016 (1036)&\hspace{-0.2 cm} 0.021 (180)\\
  \hspace{-0.2 cm} 19-20&\hspace{-0.2 cm} 0.072 (501)&\hspace{-0.2 cm} 0.028 (676)&\hspace{-0.2 cm} 0.029 (1008)&\hspace{-0.2 cm} 0.030 (109)&\hspace{-0.2 cm} 0.077 (4)\\
  \hspace{-0.2 cm} 20-21&\hspace{-0.2 cm} 0.173 (222)&\hspace{-0.2 cm} 0.063 (1080)&\hspace{-0.2 cm} 0.058 (136)&\hspace{-0.2 cm} 0.074 (2)&\hspace{-0.2 cm} -\\
  \hspace{-0.2 cm} 21-22&\hspace{-0.2 cm} 0.404 (8)&\hspace{-0.2 cm}  0.121 (394)&\hspace{-0.2 cm} 0.151 (2)&\hspace{-0.2 cm} -&\hspace{-0.2 cm} -\\
  \hline  
  \hspace{-0.2 cm} &\multicolumn{5}{c}{Tr 7}\\
  \cmidrule (lr){2-6}
  \hspace{-0.2 cm} 10-11&\hspace{-0.2 cm} 0.003 (5)&\hspace{-0.2 cm} 0.010 (3)&\hspace{-0.2 cm} 0.008 (5)&\hspace{-0.2 cm} -&\hspace{-0.2 cm} 0.005 (10)\\
  \hspace{-0.2 cm} 11-12&\hspace{-0.2 cm} 0.004 (12)&\hspace{-0.2 cm} 0.005 (15)&\hspace{-0.2 cm} 0.005 (15)&\hspace{-0.2 cm} -&\hspace{-0.2 cm} 0.004 (27)\\
  \hspace{-0.2 cm} 12-13&\hspace{-0.2 cm} 0.007 (12)&\hspace{-0.2 cm} 0.009 (12)&\hspace{-0.2 cm} 0.007 (24)&\hspace{-0.2 cm} -&\hspace{-0.2 cm} 0.005 (67)\\
  \hspace{-0.2 cm} 13-14&\hspace{-0.2 cm} 0.006 (31)&\hspace{-0.2 cm} 0.008 (40)&\hspace{-0.2 cm} 0.005 (69)&\hspace{-0.2 cm} -&\hspace{-0.2 cm} 0.006 (120)\\
  \hspace{-0.2 cm} 14-15&\hspace{-0.2 cm} 0.007 (47)&\hspace{-0.2 cm} 0.009 (67)&\hspace{-0.2 cm} 0.007 (112)&\hspace{-0.2 cm} -&\hspace{-0.2 cm} 0.007 (231)\\
  \hspace{-0.2 cm} 15-16&\hspace{-0.2 cm} 0.011 (96)&\hspace{-0.2 cm} 0.007 (116)&\hspace{-0.2 cm} 0.006 (181)&\hspace{-0.2 cm} -&\hspace{-0.2 cm} 0.008 (409)\\
  \hspace{-0.2 cm} 16-17&\hspace{-0.2 cm} 0.019 (144)&\hspace{-0.2 cm} 0.011 (170)&\hspace{-0.2 cm} 0.008 (312)&\hspace{-0.2 cm} -&\hspace{-0.2 cm} 0.013 (596)\\
  \hspace{-0.2 cm} 17-18&\hspace{-0.2 cm} 0.054 (228)&\hspace{-0.2 cm} 0.018 (280)&\hspace{-0.2 cm} 0.013 (477)&\hspace{-0.2 cm} -&\hspace{-0.2 cm} 0.021 (416)\\
  \hspace{-0.2 cm} 18-19&\hspace{-0.2 cm} 0.109 (179)&\hspace{-0.2 cm} 0.036 (441)&\hspace{-0.2 cm} 0.024 (594)&\hspace{-0.2 cm} -&\hspace{-0.2 cm} 0.051 (28)\\
  \hspace{-0.2 cm} 19-20&\hspace{-0.2 cm} 0.274 (9)&\hspace{-0.2 cm} 0.077 (598)&\hspace{-0.2 cm} 0.044 (121)&\hspace{-0.2 cm} -&\hspace{-0.2 cm} -\\
  \hspace{-0.2 cm} 20-21&\hspace{-0.2 cm} -&\hspace{-0.2 cm} 0.170 (168)&\hspace{-0.2 cm} 0.098 (7)&\hspace{-0.2 cm} -&\hspace{-0.2 cm} -\\
  \hspace{-0.2 cm} 21-22&\hspace{-0.2 cm} -&\hspace{-0.2 cm} 0.400 (7)&\hspace{-0.2 cm} -&\hspace{-0.2 cm} -&\hspace{-0.2 cm} -\\
  \hline
  \end{tabular}
\end{table}

%% file: table02.tex
\begin{table}
  \centering
  \caption{The structural parameters values obtained by us.}
  \label{struc_par}
  \begin{tabular}{c c c c c c c c c}  
  \hline  
  Cluster &  \multicolumn{2}{c}{Central coordinates} & r$_{c}$    & r$_{cluster}$ & R$_{t}$ \\
          &  RA (J2000)      & DEC (J2000)           & ($\prime$) &   ($\prime$)  & (pc)   \\ \hline
   IC 1442 &22:16:03.7&+53:59:29.4&1.7$\pm$0.5&9.5$\pm$0.5&10.70\\
   King 21 &23:49:55.0&+62:42:18.0&1.0$\pm$0.2&8.2$\pm$0.5&10.64\\
   Tr 7    &07:27:23.8&-23:56:56.4&1.0$\pm$0.3&7.1$\pm$0.5&7.82\\ \hline
  \end{tabular}
\end{table}

%% file: table03.tex
\begin{table}
  \centering
  \caption{The mean proper motions, in mas yr$^{-1}$ units, obtained for IC 1442, King 21, and Tr 7.}
  \label{pm}
  \hbox{ 
  \begin{tabular}{c c c c c c c c c c c}
  \hline
     & IC 1442 & King 21 & Tr 7 \\ \hline
 \underline{Member stars:} &  \\
   $\bar{\mu}_{\alpha*}$& -3.38$\pm$0.25& -3.24$\pm$0.23& -2.49$\pm$0.14 \\
  $\bar{\mu}_{\delta}$ & -3.21$\pm$0.22& -1.64$\pm$0.22& 3.34$\pm$0.17 \\
  $\bar{\mu}$ & 4.67$\pm$0.23& 3.64$\pm$0.22& 4.17$\pm$0.15 \\
\underline{Field stars:} & \\
$\bar{\mu}_{\alpha*}$ & -2.39$\pm$4.01&-1.72$\pm$4.47& -1.50$\pm$2.80 \\
  $\bar{\mu}_{\delta}$ & -2.45$\pm$3.13&-0.96$\pm$2.68& 2.01$\pm$ 3.90 \\ \hline
  \end{tabular}
}
\end{table}

%% file: table04.tex
\begin{table} 
  \centering
  \caption{The slopes of the (V-$\lambda$)/(B-V) diagrams for the member stars and corresponding normal values}.
  \label{rv_slope}
  \hbox{
\hspace{-0.4 cm}  
  \begin{tabular}{c c c c c c c c}  
  \hline  
 \hspace{-0.3 cm} & \hspace{-0.3 cm} $\frac{(V-R)}{(B-V)}$ & \hspace{-0.3 cm} $\frac{(V-I)}{(B-V)}$ & \hspace{-0.3 cm} $\frac{(V-J)}{(B-V)}$ & \hspace{-0.3 cm} $\frac{(V-H)}{(B-V)}$ & \hspace{-0.3 cm} $\frac{(V-K)}{(B-V)}$ \\ \hline
 \hspace{-0.3 cm} IC 1442  & \hspace{-0.3 cm} 0.75$\pm$0.02 & \hspace{-0.3 cm} 1.28$\pm$0.03 & \hspace{-0.3 cm} 2.21$\pm$0.04 & \hspace{-0.3 cm} 2.89$\pm$0.08 & \hspace{-0.3 cm} 3.08$\pm$0.10 \\
 \hspace{-0.3 cm}  King 21  & \hspace{-0.3 cm} 0.64$\pm$0.02 & \hspace{-0.3 cm} 1.19$\pm$0.03 & \hspace{-0.3 cm} 2.06$\pm$0.04 & \hspace{-0.3 cm} 2.63$\pm$0.06 & \hspace{-0.3 cm} 2.76$\pm$0.07 \\
  \hspace{-0.3 cm}  Tr 7 & \hspace{-0.3 cm} - & \hspace{-0.3 cm} 1.21$\pm$0.02 & \hspace{-0.3 cm} 1.94$\pm$0.04 & \hspace{-0.3 cm} 2.54$\pm$0.04 & \hspace{-0.3 cm} 2.88$\pm$0.09 \\
  \hspace{-0.3 cm} Normal & (0.55) & (1.10) & (1.96) & (2.42) & (2.60) \\ \hline
  \end{tabular}
  }
\end{table}

%% file: table05.tex
\begin{table*}
  \centering
  \caption{The values of physical parameters obtained in the present study.}
  \label{clust_par}
  \begin{tabular}{c c c c c c c c c c}  
  \hline  
   \hspace{-0.2 cm} Cluster & \hspace{-0.3 cm} r$_{c}$ & r & \hspace{-0.3 cm} $\bar{\mu}$ &R$_{cluster}$ & \hspace{-0.3 cm} E(B-V) & \hspace{-0.3 cm} log(Age)& distance\\
    & \hspace{-0.3 cm} (pc) & (pc)& \hspace{-0.3 cm} (mas/yr)& &(mag) & \hspace{-0.3 cm} (yr) & \hspace{-0.3 cm} (pc) \\ \hline
  IC 1442 &1.4$\pm$0.6&7.9$\pm$1.1&4.67$\pm$0.23&3.8&0.54$\pm$0.04&7.40$\pm$0.30&2847$\pm$238\\
  King 21 &0.8$\pm$0.2&6.6$\pm$0.8&3.64$\pm$0.22&3.4&0.76$\pm$0.06&7.70$\pm$0.20&2622$\pm$156\\
  Tr 7 &0.4$\pm$0.2&3.2$\pm$0.4&4.17$\pm$0.15&3.3&0.38$\pm$ 0.04&7.85$\pm$0.25&1561$\pm$74\\ \hline
  \end{tabular}
\end{table*}

%% file: table06.tex
\begin{table*}
  \centering
  \caption{The LF and MF values derived for IC 1442, King 21, and Tr 7.}
  \label{lf}
  \begin{tabular}{c c c c c c c c c c c c c c}
  \hline
   \hspace{0.3 cm} V \hspace{0.3 cm}& \multicolumn{3}{c}{IC 1442} \hspace{0.3 cm}& \multicolumn{3}{c}{King 21} \hspace{0.3 cm}&\multicolumn{3}{c}{Tr 7} \\
   \cmidrule(lr){2-4}\cmidrule(lr){5-7}\cmidrule(lr){8-10}
   Range & \hspace{0.3 cm} Mass Range& \hspace{0.3 cm}$\bar{m}$& \hspace{0.3 cm} N& \hspace{0.3 cm} Mass Range& \hspace{0.3 cm} $\bar{m}$& \hspace{0.3 cm} N& \hspace{0.3 cm} Mass Range& \hspace{0.3 cm} $\bar{m}$& \hspace{0.3 cm} N \\
 (mag)& \hspace{0.3 cm} (M$_{\odot}$) & \hspace{0.3 cm} (M$_{\odot}$) & \hspace{0.3 cm}& \hspace{0.3 cm} (M$_{\odot}$) & \hspace{0.3 cm} (M$_{\odot}$) & \hspace{0.3 cm}& \hspace{0.3 cm} (M$_{\odot}$) & \hspace{0.3 cm} (M$_{\odot}$) \\
    \hline
    10-11& \hspace{0.3 cm} -& \hspace{0.3 cm}-& \hspace{0.3 cm} -& \hspace{0.3 cm} 6.891-6.889& \hspace{0.3 cm} 6.890& \hspace{0.3 cm} 1& \hspace{0.3 cm} -& \hspace{0.3 cm} -& \hspace{0.3 cm}- \\ 
    11-12& \hspace{0.3 cm} 9.31-7.99& \hspace{0.3 cm} 8.21& \hspace{0.3 cm} 2& \hspace{0.3 cm} 6.89-6.70& \hspace{0.3 cm} 6.89& \hspace{0.3 cm} 1& \hspace{0.3 cm} 4.96-3.87& \hspace{0.3 cm} 4.94& \hspace{0.3 cm} 1 \\    
    12-13& \hspace{0.3 cm} 7.99-6.24& \hspace{0.3 cm} 6.77& \hspace{0.3 cm} 3& \hspace{0.3 cm} 6.70-5.89& \hspace{0.3 cm} 6.47& \hspace{0.3 cm} 1& \hspace{0.3 cm} 3.87-2.80& \hspace{0.3 cm} 3.19& \hspace{0.3 cm} 8 \\
    13-14& \hspace{0.3 cm} 6.24-4.36& \hspace{0.3 cm} 5.37& \hspace{0.3 cm} 5& \hspace{0.3 cm} 5.89-4.62& \hspace{0.3 cm} 5.21& \hspace{0.3 cm} 12& \hspace{0.3 cm} 2.80-2.00& \hspace{0.3 cm} 2.35& \hspace{0.3 cm} 15 \\
    14-15& \hspace{0.3 cm} 4.36-3.01& \hspace{0.3 cm} 3.65& \hspace{0.3 cm} 18& \hspace{0.3 cm} 4.62-3.34& \hspace{0.3 cm} 3.92& \hspace{0.3 cm} 16& \hspace{0.3 cm} 2.00-1.56& \hspace{0.3 cm} 1.79& \hspace{0.3 cm} 15\\
    15-16& \hspace{0.3 cm} 3.01-2.09& \hspace{0.3 cm} 2.46& \hspace{0.3 cm} 33& \hspace{0.3 cm} 3.34-2.35& \hspace{0.3 cm} 2.74& \hspace{0.3 cm} 21& \hspace{0.3 cm} 1.56-1.28& \hspace{0.3 cm} 1.43& \hspace{0.3 cm} 21\\
    16-17& \hspace{0.3 cm} 2.09-1.59& \hspace{0.3 cm} 1.83& \hspace{0.3 cm} 36& \hspace{0.3 cm} 2.35-1.72& \hspace{0.3 cm} 1.97& \hspace{0.3 cm} 38& \hspace{0.3 cm} 1.28-1.07& \hspace{0.3 cm} 1.18& \hspace{0.3 cm} 23\\
    17-18& \hspace{0.3 cm} 1.59-1.32& \hspace{0.3 cm} 1.45& \hspace{0.3 cm} 66& \hspace{0.3 cm} 1.72-1.39& \hspace{0.3 cm} 1.54& \hspace{0.3 cm} 49& \hspace{0.3 cm} 1.07-0.90& \hspace{0.3 cm} 1.02& \hspace{0.3 cm} 20\\
    18-19& \hspace{0.3 cm} -& \hspace{0.3 cm} -& \hspace{0.3 cm} -& \hspace{0.3 cm} 1.39-1.16& \hspace{0.3 cm} 1.28& \hspace{0.3 cm} 47& \hspace{0.3 cm} 0.90-0.76& \hspace{0.3 cm} 0.82& \hspace{0.3 cm} 14\\
 \hline
  \end{tabular}
\end{table*}

%% file: table07.tex
\begin{table}\tiny 
  \centering
  \caption{The MF slopes for IC 1442, King 21, and Tr 7. The inner region slopes are obtained in circular regions having radii equal to 6$^{\prime}$.5, 4$^{\prime}$.0, and 4$^{\prime}$.0 for IC 1442, King 21, and Tr 7, respectively.}
  \label{mf_slope}
  \begin{tabular}{c c c c c c c c c}  
  \hline  

 &\multicolumn{2}{c}{entire}&\multicolumn{2}{c}{inner}&\multicolumn{2}{c}{outer} \\ 
 \cmidrule(lr){2-3}\cmidrule(lr){4-5}\cmidrule(lr){6-7}
 Cluster& Range& {$\Gamma$}& Range& {$\Gamma$}& Range& {$\Gamma$}\\
        & (M$_{\odot}$)&   & (M$_{\odot}$)&   & (M$_{\odot}$)& \\
 \hline 
 IC 1442&9.31-1.32&-1.94$\pm$0.18&9.31-1.32&-1.64$\pm$0.14&9.31-1.32&-1.84$\pm$0.24 \\
 King 21&6.89-1.16&-1.54$\pm$0.32&6.89-1.16&-1.06$\pm$0.32&5.89-1.16&-1.93$\pm$0.15 \\
 Tr 7&4.96-1.07&-2.31$\pm$0.29&4.96-1.07&-1.78$\pm$0.32&3.87-1.07&-2.37$\pm$0.81 \\

\hline
  \end{tabular}
\end{table}

%% file: table08.tex
\begin{table}
\centering
\caption{The dynamical parameters calculated for IC 1442, King 21, and Tr 7.}
\label{dyna}
\begin{tabular}{lccc}
\hline
Cluster parameter                     & IC 1442 & King 21 & Tr 7 \\ \hline
Number of member stars used           & 257& 236& 124   \\
Mean stellar mass ($\bar{m}/M_{\odot}$) & 1.88& 1.96& 1.50 \\
Total mass  ($M_\odot$)               & 483.16& 462.56& 186.00 \\
Cluster half radius ($r_h/pc$)        & 4.9& 2.3& 1.7   \\
Tidal radius ($r_t/pc$)               & 10.70& 10.64& 7.82  \\
Relaxation time ($T_E/Myrs$)          & 74& 26& 34 \\ \hline
\end{tabular}
\end{table}